\documentclass[aps,reprint,amsmath,amssymb,superscriptaddress,floatfix,twocolumn]{revtex4-2}
\usepackage{amsmath}
\usepackage{changepage}
\usepackage{epsfig}
\usepackage{epstopdf}
\usepackage[colorlinks,citecolor=blue,linkcolor=blue]{hyperref}
\usepackage{lipsum}
\usepackage{booktabs}
\usepackage{multirow}
\usepackage{braket} 
\usepackage{multirow}
\usepackage{makecell}
\usepackage{amsmath}
\usepackage{mathtools}

\begin{document}
\title{Realizing the controllable excitation transfer based on the  atom coupling the finite-size Su-Schrieffer-Heeger model}
\author{Da-Wei Wang}
\affiliation{School of Physics, Dalian University of Technology, Dalian 116024,People's Republic of China}
\author{Chengsong Zhao}
\affiliation{School of Physics, Dalian University of Technology, Dalian 116024,People's Republic of China}
\author{Junya Yang}
\affiliation{School of Physics, Dalian University of Technology, Dalian 116024,People's Republic of China}
\author{Ye-Ting Yan}
\affiliation{School of Physics, Dalian University of Technology, Dalian 116024,People's Republic of China}
\author{Ling Zhou}
\affiliation{School of Physics, Dalian University of Technology, Dalian 116024,People's Republic of China}
\begin{abstract}
  In this paper, we study the interaction between  atom  and the finite-size Su-Schrieffer-Heeger (SSH) model.  We find that when the finite SSH model in the trivial phase, it can be viewed as the  atom coupling with the  waveguide with the finite bandwidths and non-linear dispersion relation. However,  for the SSH model in the topological phase,  when we consider the frequency of the atom is resonant with the edge mode of the SSH model, we find that the atom state couples to the two edge states. In this case, we find that there exists a special channel that can be utilized to transfer the atomic excitation to the ends of the SSH model using adiabatic processes. When the atom couples to the different sub-lattice, the excitation of the atom can be transferred to the leftmost  or rightmost end of the chain, which provides the potential application toward quantum information processing. Furthermore, The excitation transfer of excited states of atoms to the ends of the chain can also be realized without the adiabatic process. Our work provides a pathway for realizing controllable quantum information transfer based on the atom coupled topological matter.
\end{abstract}

\maketitle

\section{Introduction}
Topological insulators have attracted much interest and attention in quantum physics due to their many interesting properties, including robustness to local decoherence processes and potential applications in quantum information \cite{RevModPhys.82.3045, Wray2010,Shalaev2019,RevModPhys.91.015006}.
The chiral edge states of topological photonic systems yield directional transport of photons and phonons, which has been exploited in the design of amplifiers  and topological lasers \cite{Goblot2017,PhysRevLett.120.113901,Zhang2022,PhysRevA.103.033513}.  The transport phenomena in dissipative systems, which are characterized by topological winding numbers, have been studied in \cite{Wanjura2020,PhysRevLett.122.143901}.
The Su-Schrieffer-Heeger (SSH) model was originally used to describe the transport properties of conducting polyacetylene \cite{PhysRevLett.42.1698,RevModPhys.60.781}, it has attracted increasing attention as the simplest topological insulator model with a simple structure and rich physical pictures  \cite{RevModPhys.88.021004,10.3389/fphy.2021.813801, Meier2016,Wanjura2020}.  The SSH model and its various extensions have been used in quantum information processes for controlled quantum state transfer \cite{PhysRevLett.129.215901,PhysRevA.103.052409,PhysRevA.98.012331,PhysRevA.106.052411,PhysRevA.102.022404,PhysRevA.103.023504},as well as in topological transmission devices such as topological beam splitter, topological router \cite{PhysRevApplied.18.054037,PhysRevA.107.062214,PhysRevResearch.3.023037,PhysRevB.103.085129}.

Waveguide quantum electrodynamics involving the coupling of atoms and one-dimensional propagation fields has become an important physical platform for realizing quantum information processing  \cite{Brehm2021,Zanner2022,goban2014atom,PhysRevLett.115.063601,RevModPhys.89.021001,PhysRevLett.111.053601,doi:10.1126/science.ade7651,doi:10.1126/science.ade9324,doi:10.1126/science.1244324,RevModPhys.95.015002} and quantum simulation  \cite{Douglas2015, RevModPhys.86.153,PhysRevX.11.011015,Wang:22}. When the transition frequencies of atoms lie in the photonic band gap, photons in waveguides are localized around them, which forms bound states that can mediate coherent interactions between atoms \cite{Liu2017,PhysRevLett.123.233602,PhysRevLett.64.2418,PhysRevX.6.021027}.
This yields many interesting phenomena and applications such as the generation of long-range entanglement \cite{PhysRevLett.110.080502,PhysRevLett.106.020501}, photon transport \cite{PhysRevX.5.041036,PhysRevA.82.063816}, unconventional quantum optics \cite{Bello2019, doi:10.1126/science.ade9324,PhysRevLett.126.203601,PhysRevA.104.053522,PhysRevResearch.4.023077} and the simulation of topological states \cite{Ringel_2014,PhysRevB.105.094422,PhysRevA.82.063816}. When the atoms couple to the topological waveguide under the periodic boundary, both the atom and the photon behave in exotic ways, for example, the  atom can view as an effective boundary and induce the chiral zero-energy modes exhibiting a distribution of chirality as well as the robustness to the off-diagonal disorder \cite{PhysRevX.11.011015,Bello2019,PhysRevA.106.033522}.


However, few attention has been paid to the coupling of atom to a finite-size topological waveguides. In this paper, we consider the atom couples to the finite SSH model. We find that when the atom couples to the finite SSH model within the bandgap in the trivial phase, a stable bound state is formed consisting of atom in the excited state and localized waveguide photons.However, in the topological phase, we find that when the atomic frequency is not resonant with the SSH model but in the bandgap, the atom-photons bound state is formed, while if the atomic frequency is nearly resonant with the center frequency of the 
SSH model, the atom-photons bound state vanishes. Under this case, we find the atom will couple to edge states and there is a special channel allowing the transfer of excitations of atom to both ends of the chain by adiabatically adjusting parameter. Furthermore, we find that the transfer of information to the leftmost and rightmost ends of the chain depends on the coupling point of atom to the sub-lattice. Moreover, we can also realize the transfer of excitation of atom to the left or right edges states without the adiabatic process.

\section{atom couples the finite SSH model}

\begin{figure}[t]
  \centering
  \includegraphics[width=6cm]{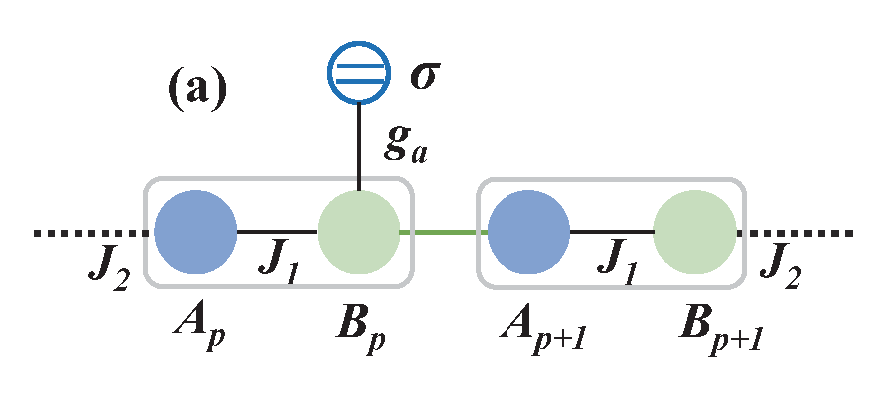}
  \caption{(a)Schematic illustration of the atom couples to the finite SSH models, where the  atom is coupled to the sub-lattice $A$ at the $q$th cell or to the sub-lattice $B$ at the $p$th cell with the coupling strength $g$.}
  \label{model}
\end{figure}

 We consider an atom coupling to the finite SSH model, where the atom is coupled to the sub-lattice $A$ or the sub-lattice $B$ with the coupling strength $g$, as shown in Fig. \ref{model}. The total Hamiltonian can be written as
\begin{equation}
  \begin{aligned}
   H&=H_{SSH}+H_I+\omega_e \sigma^\dag\sigma,
  \end{aligned}\label{ee1}
  \end{equation}
with
\begin{subequations}
  \begin{align}\label{eq1}
    H_{\mathrm{SSH}}&=\sum_{i=1}^N\omega_o(a_{i}^{\dagger}a_{i}+b_{i}^{\dagger}b_{i}) +(J_1a_{i}^{\dagger}b_i+ J_2a_{i+1}^{\dagger}b_i+h.c.),\\
    H_I&=(gc_{q}^{\dagger}\sigma+h.c.),
  \end{align}
\end{subequations}
where $H_{\mathrm{SSH}}$ is the Hamiltonian of the SSH model, which is a lattice model  describes one-dimensional hopping with period of 2. Each cell of the lattice contains two sub-lattices $A$ and $B$ with frequencies $\omega_o$. The inter-cell and extra-cell and coupling are $J_1=J(1+\cos\theta)$ and $J_2=J(1-\cos\theta)$  with $\theta \in [0,2\pi]$, respectively.
Strictly speaking, the free terms in $H_{\mathrm{SSH}}$ partly breaks chiral symmetry of a standard SSH model \cite{RevModPhys.91.015006,RevModPhys.88.021004}, but the quantized Zak phase remains as long as the frequency of each site is not perturbed \cite{PhysRevB.84.195452,PhysRevLett.62.2747}. Then, we will use the frequency $\omega_o$ as the reference energy in the following discussion. $H_I$ represent the interaction between the atom and the SSH model, where $\{ c_{q(p)}={a_q, b_p}\}$.

In the single-excited space, we can consider $|A_i\rangle=a_i^\dag|G\rangle$ and $|B_i\rangle=b_i^\dag|G\rangle$, where $|G\rangle$ is the ground state of the SSH model.
The Hamiltonian (\ref{eq1}) in single-excited space can be written as $H_{\rm SSH}=\sum_{i=1}^N \omega_o(|A_i\rangle \langle A_i| |B_i \rangle \langle B_i|) + ( J_1|A_i\rangle\langle B_i|+ J_2|A_{i+1}\rangle\langle B_i|+h.c.)$,
We can diagonalize the Hamiltonian as $H_{SSH}=\sum_{j=1}^{2N}E_j|\Psi_j\rangle\langle\Psi_j|$, where $E_j$ and $|\Psi_j\rangle$ are the eigenenergies and Eigenvectors. The Eigenvectors $|\Psi_j\rangle$ can be written as the superposition of the $|A_i\rangle$ and $|B_i\rangle$ as $|\Psi_j\rangle=\sum_{i=1}^N(\zeta_{2i-1,j}|A_i\rangle+\zeta_{2i,j}|B_i\rangle)$, where $\zeta_{2i-1,j}=\langle A_i|\Psi_j\rangle$ and $\zeta_{2i,j}=\langle B_i|\Psi_j\rangle$ are the projections of the eigenvectors on the sub-lattice $A_i, B_i$. Conversely the basis $|A_i\rangle$ and $|B_i\rangle$ can also be written as the superposition of the eigenvectors as $|A_i\rangle=\sum_{j=1}^N\zeta_{2i-1,j}^*|\Psi_j\rangle$ and $|B_i\rangle=\sum_{j=1}^N\zeta_{2i,j}^*|\Psi_j\rangle$.

\begin{figure}[t]
  \centering
  \includegraphics[width=9cm]{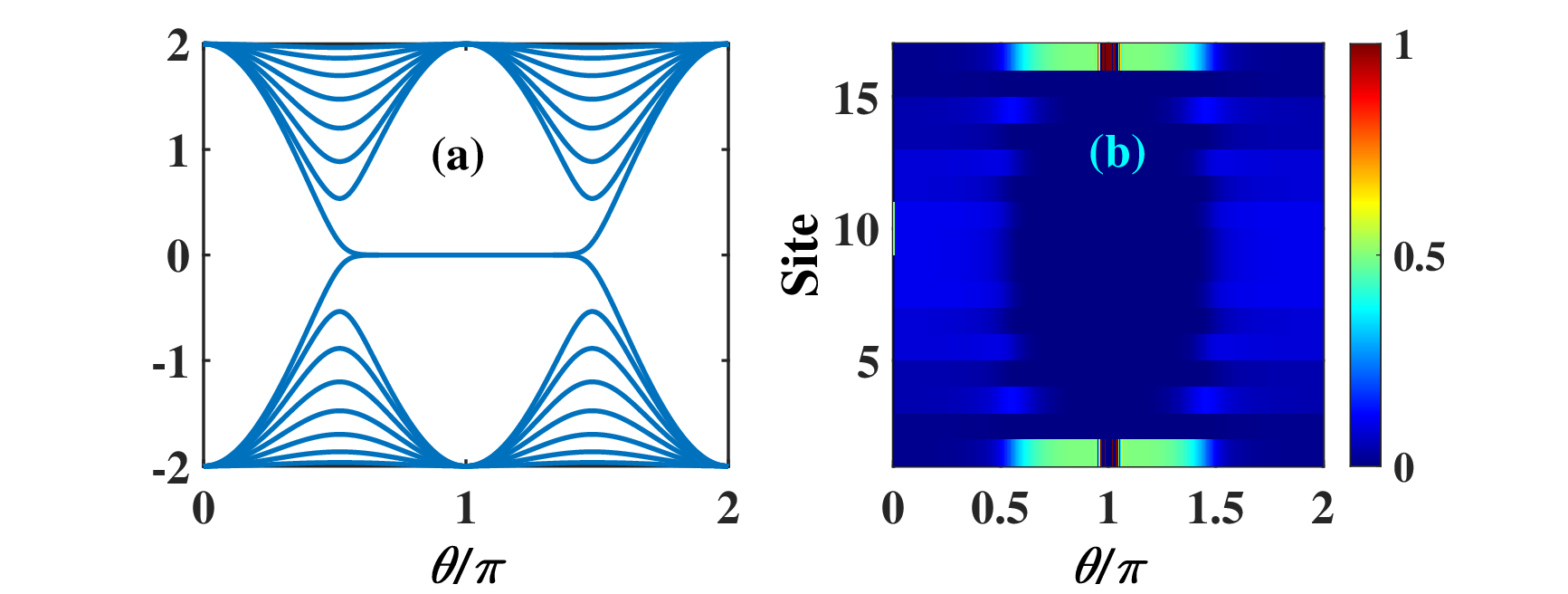}
  \caption{(a) The energy spectrum of SSH model varies with the variation of $\theta$. (b) The probability distribution of the eigenstates corresponding to (a). The parameters are $N=8$ and $J=1$}.\label{Nojyz}
 \end{figure}

In the topological phase of SSH model, there are two bandgap states, and we denote them as $|\Psi_{N(N+1)}\rangle$ with energy $E_{N(N+1)}$.
In Fig. \ref{Nojyz} (a), we plot the energy spectrum varying with $\theta$ under open boundary condition. It can be seen that the SSH model possesses two bandgap states in the gap when $\theta \in [0.5\pi, 1.5\pi]$ (topological phase). The probability distribution of the bandgap states varying with $\theta$ is shown in Fig. \ref{Nojyz} (b). It is clear that the band gap states is mainly located at the two ends of the chain with the same probability. This phenomenon originates from the size effect.

For the larger lattice $N\to\infty$, Ww can analytically derive the distribution of {the bandgap states as
\begin{equation}
  \begin{aligned}
    |\psi_L\rangle=N_L\sum_{i=1}^N c_i^a|A_i\rangle, \quad
    |\psi_R\rangle=N_R\sum_{i=1}^N c_i^b|B_i\rangle. \label{ed}
  \end{aligned}
\end{equation}
where $c_i^a=(-\frac{J_1}{J_2})^{i-1}$, $c_i^b=(-\frac{J_1}{J_2})^{N-i}$, $N_L=N_R=[1-(J_1/J_2)^2]^{1/2}[1-(J_1/J_2)^{2N}]^{-1/2}$ are the probability amplitude on the sub-lattice $A$, $B$ and  the normalization factors, respectively.
For short lattices, the edge states are not separated but hybridized to each other. The  energy can also be obtain $E_{N(N+1)}=\pm(-1)^{N+1}N_L^2J_1(J_1/J_2)^{N-1}$ and the corresponding hybridized edge states can be written as
\begin{equation}
   |\Psi_{N(N+1)}\rangle =\frac{1}{\sqrt{2}}(|\psi_L\rangle\pm|\psi_R\rangle)
\end{equation}
Therefore, we can obtain $\zeta_{2i-1,N(N+1)}=N_Lc_i^a$ and $\zeta_{2i,N(N+1)}=N_Rc_i^b$.
When the atom couples the sub-lattice $A$, the total Hamiltonian can be rewritten as
\begin{equation}
  \begin{aligned}
   H&=\Delta \sigma^\dag\sigma+\sum\limits_{j=1}^{2N}[E_j|\Psi_j\rangle\langle\Psi_j| +(g\zeta^*_{2q-1,j}\Psi_j^\dag\sigma+h.c.)], \label{e5}
\end{aligned}
\end{equation}
where  $\Psi_j^\dag=|\Psi_j\rangle\langle G|$.
We consider that the atomic frequency is resonant with $\omega_o$, that is $\Delta \approx 0$ and the coupling between atom and the lattice is weak $g/J \ll 1$.
In Figs. \ref{detuning_energy_evolution} (a), we plot the energy spectrum varying with $\Delta$ for the chain in the trivial phases.
We find that in the trivial phase, due to the large detuning from the frequency of atom $|E_j-\Delta| \gg g$, the atom will decouple from the chain. From the viewpoint of dynamical evolution, a bound state is formed consisting of atom in the excited state and waveguide photons, where the waveguide photon distribution is exponentially localized around the atom, thus the emission of photon from the atom to the SSH model is prohibited. In a word, in the trivial phase, the atom decouples with the SSH lattice.

In the topological phase, there are two edge state with $E_{N(N+1)} \thicksim \sum_{j=1}^{2N} g\zeta_{2q-1,j} \approx 0$ which is near resonant with the atom Thus the effective coupling between the edge states the atom may be obtained. In Fig. \ref{detuning_energy_evolution}(b) and (c), we plot the spectrum of the system when the SSH lattice is in the topological phase. We find that when $\Delta = 0.1J$, the eigenenergies of the system domain by left and right edge state as well as the atom, separately. As shown in Figs. \ref{detuning_energy_evolution} (c), the effective exchange between the atom and the edge states can not be observed. When $\Delta =0.04J$, the three hybridized eigenmodes composed of edge states and the atom can be observed from Fig. \ref{detuning_energy_evolution}(c). Meanwhile, in Fig. \ref{detuning_energy_evolution}(d), we observe the Rabi oscillation between the atom and the two edge states, which means the effective coupling. In Figs. \ref{detuning_energy_evolution} (f) and (g), we plot the spatial distribution of the eigenstates corresponding to the purple and red energy levels, and we can see that this distribution is exactly the spatial distribution of the hybridized states.
The the coupling between other eigenstates and atom can be ignored due to $|E_j|_{Mix}=|E_{N-1(N+2)}|\gg \sum_{j=1}^{2N} g\zeta^*_{2q-1,j}$. Thus the effective Hamiltonian can be written as
\begin{equation}
  \begin{aligned}
   H&=E_N\Psi_{N}^\dag\Psi_{N}+E_{N+1}\Psi_{N+1}^\dag\Psi_{N+1}\\
   \quad &+ N_Lgc_q^a/\sqrt{2}(\Psi_{N}^\dag+\Psi_{N+1}^\dag)\sigma + h.c., \label{e6}
\end{aligned}
\end{equation}
Therefore, in single-excitation space, when we consider the frequency of the  atom is resonant with the SSH model, the above Hamiltonian  only causes the following subspace transitions $|\psi_N,g\rangle\leftrightarrow |vac, e\rangle$ $\leftrightarrow |\psi_{N+1},g\rangle$. And the case of the atom coupling to sub-lattice B can also be analyzed similarly, we do not go into detail.

\begin{figure}[t]
  \centering
  \includegraphics[width=9cm]{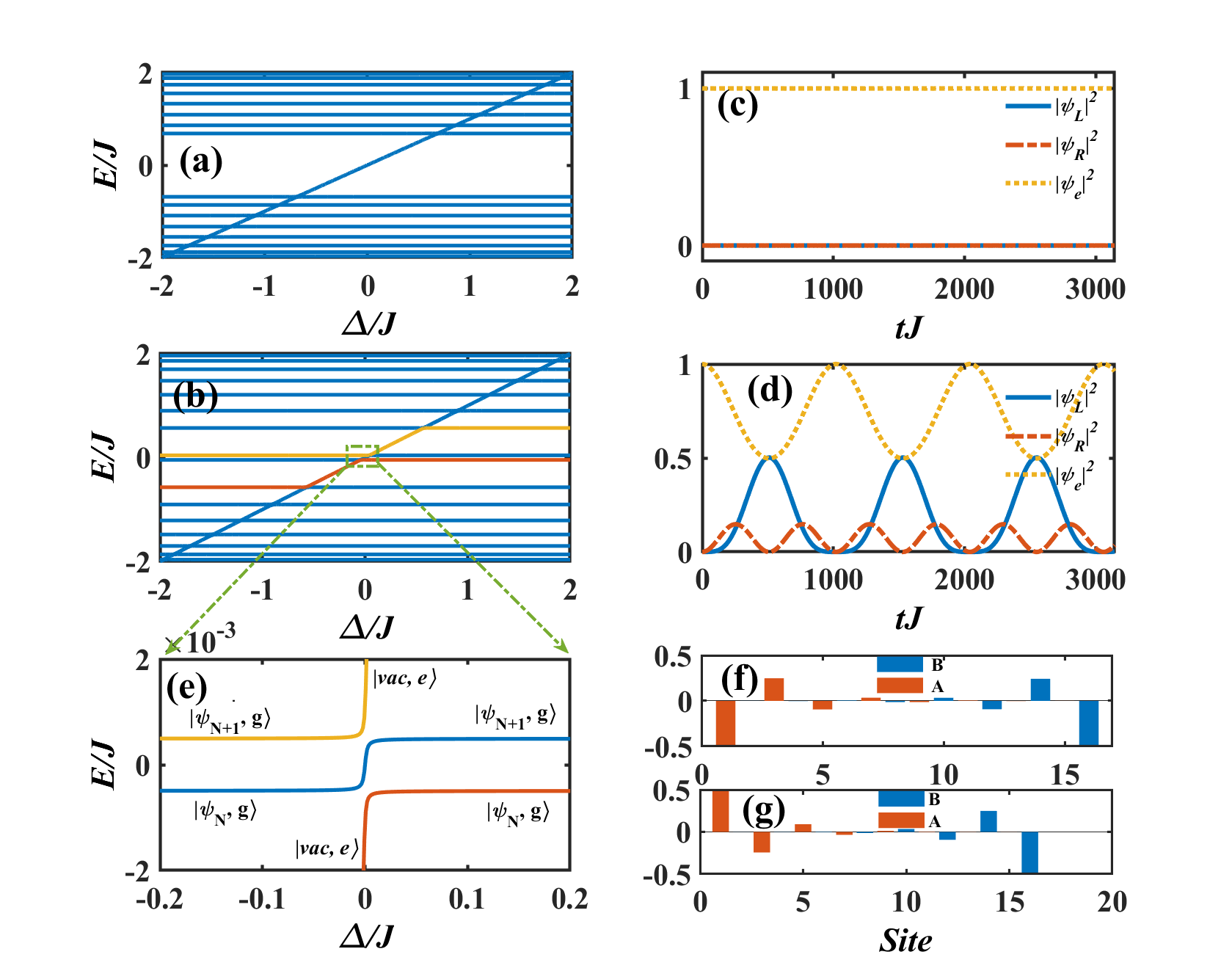}
  \caption{(a) (b) The energy spectrum versus $\Delta$ for the SSH model in the trivial $\theta=0.4\pi$ and topological phases $\theta=0.65\pi$.
  (c) (d) The evolution of the excited state of the atom $\psi_e$ and the left edge state $|\psi_{L}\rangle$ and the right edge state $|\psi_{R}\rangle$ with $\Delta=0.1J, 0$ for $0.65\pi$, respectively. (e) The detail  of (b) at $\Delta \in [-0.05, 0.05$]. (f) and (g) Bandgap states marked in orange and blue in spatial profiles for (c) at $\Delta=0.04J$.  The other  parameters are $p=5$,  $N=8$,  $g=0.01J$ and $J=1$.}
  \label{detuning_energy_evolution}
\end{figure}

 \section{controllable excitation transfer by the adiabatic process}

When the atom couples to the sub-lattice A or B, the effective interaction between the edge states and the atom can be rewritten Eq. (\ref{e6}) in matrix form in the subspace \{$|vac, e\rangle$, $| \psi_{L}, g\rangle$, $| \psi_{R}, g\rangle$\}
 \renewcommand\arraystretch{1.5}
 \begin{equation}
H_{sub} =\left [ \begin{matrix}
 0& G_{L,a}& G_{R,b} \\
 G_{L,a}& 0& G \\
  G_{R,b}& G& 0 \\
  \end{matrix} \right ],\label{H_{eff}}
 \end{equation}
 where we set $G=E_N$, $G_{L,a}= gN_Lc_i^a (G_{L,a}=0)$ and  $G_{R,b}=0 (G_{R,b}=0gN_Lc_i^b$) for the case of the atom coupling to sub-lattice A (B).
The eigenvalues of $H_{sub}$ can be obtained by solving the cubic equation
 \begin{equation}
   \lambda^3-(G^2+G_{L,a}^2+G_{R,b}^2)\lambda+2G_{R,b}G_{L,a}G=0. \label{e8}
 \end{equation}
When the atom couples to the sub-lattice $B$, the roots of the equation (\ref{e8}) can be solved as
$\lambda_{0(\pm)} =0, \pm\sqrt{G^2+G_{R,b}^2}$. The corresponding eigenstates can be obtained $|\phi_0\rangle=N_0\{-G/G_{R,b},1,0\}$, $|\phi_\pm\rangle=N_\pm\{\pm\frac{G_{R,b}}{\sqrt{G^2+G_{R,b}^2}}, \pm\frac{G}{\sqrt{G^2+G_{R,b}^2}},1\}$, where $N_0$ and $N_\pm$ are the normalization coefficient.

In Fig. \ref{energy_p}(a), we plot the parameters $G$ and $G_{R,b}$  varying with $\theta$ for different coupling position $p$. It can be find that 
when $\theta\approx 0.73\pi$, $G=0$ and for fixed  $\theta$,  $G_{R,b}$ gradually becomes larger as the coupling position $p$ becomes larger.  We can observe that for $p<5$, $G$ is always greater than $G_{R,b}$  and for $p\geq 5$, $G$ change from greater than $G_{R,b}$ to less than $G_{R,b}$ as $\theta$ varies to $0.7\pi$.  
In Fig. \ref{energy_p}(b),
we plot the roots varying with $\theta$. We can see with increasing $\theta\to \pi$, $\lambda_{\pm}$ gradually approach zero, due to  $G=0$ ($G_{R,b}=0$) when $\theta\approx 0.73\pi$ ($0.8\pi$) for $p=6$ as shown in the  Fig. \ref{energy_p}(b).  

For $\lambda=0$, we find that $|\phi_0\rangle$ is a superposition of atom excited state and the left edge state. It can be seen that  $|\phi_0\rangle$ is distributed with probability $\frac{GG_{R,b}}{G^2_{R,b}+G^2}$ in the excited state of the atom and probability  $\frac{G^2_{R,b}}{G^2_{R,b}+G^2}$ in the left edge state.
If we slowly adiabatically tune $\theta$ from $0.5\pi\to\pi$ and consider that $G_{R,b}$ tends to zero slower than $G$, then we can realize the excitation transfer of the atom to the left edge state.  This process requires that the rate of adiabatic evolution be no greater than the spacing of neighboring energy levels $\Delta_{\lambda}=|\lambda_{\pm}-\lambda_0|$ \cite{PhysRevA.98.012331,PhysRevB.99.155150,PhysRevA.103.052409}. 
Therefore the coupling position of the atoms $p\geq5$ is necessary to satisfy that $G_{R,b}$ tends to zero more slowly than $G$ changes with $\theta$, so that the energy band still has a gap $\Delta_{\lambda}=|G_{R,b}|$ and excitation transfer is completed.
For example, when we consider  $p=6$, because the energy levels close at $\theta\approx 0.8\pi$, we have to complete the adiabatically state transfer before $\theta\approx 0.8\pi$. }
 \begin{figure}[t]
  \centering
  \includegraphics[width=9cm]{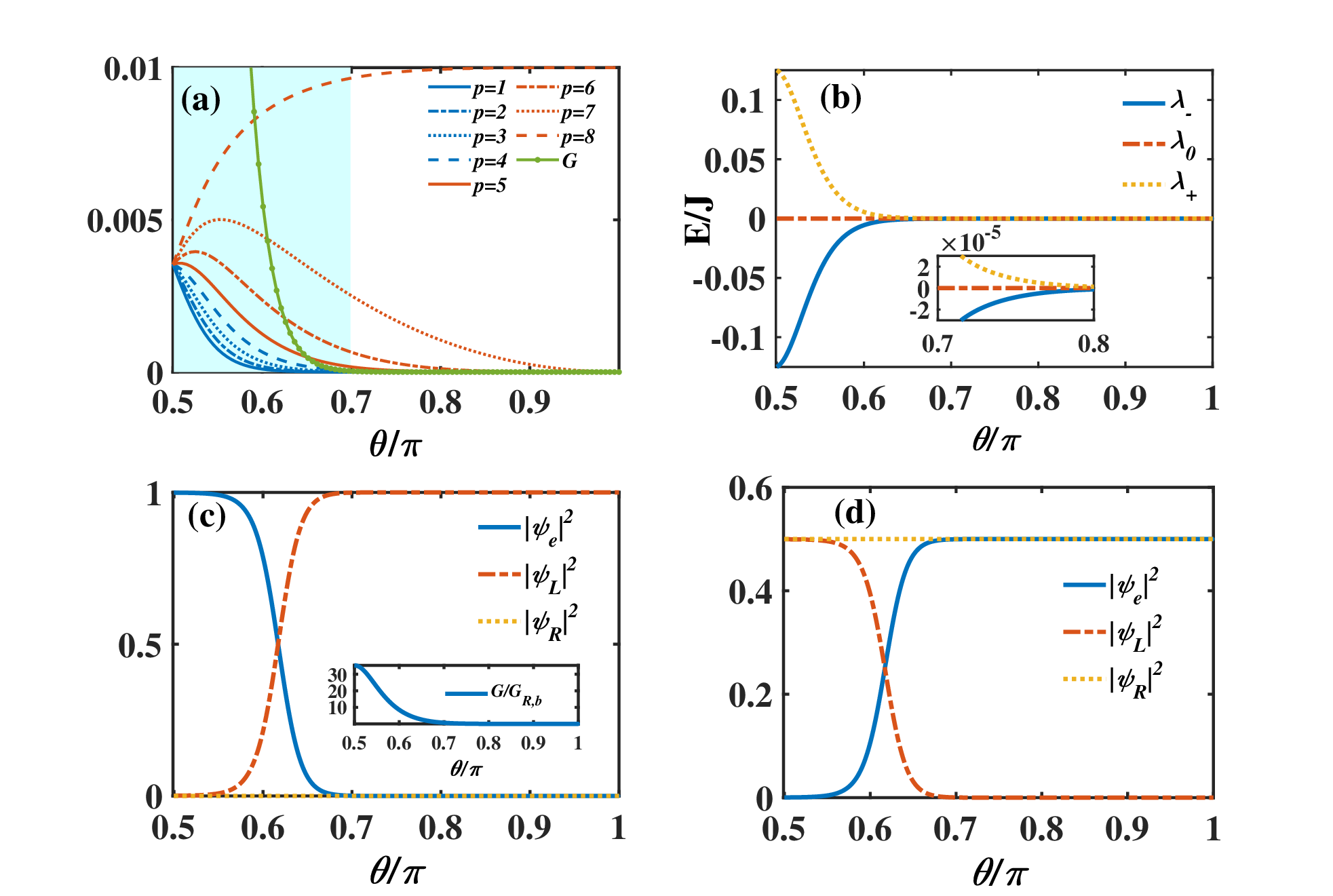}
  \caption{(a) The parameters $G$ and  $G_{R,b}$  as a function of $\theta$ for different $p$. (b) The eigenenergies vary with  $\theta$ for $p=6$. Distribution of the probability of eigenstates in excited state atoms, left and right edge states (c) $|\phi_0\rangle$, (d) $|\phi_+\rangle$  for $p=6$. The parameter are   $N=8$, $g=0.01J$ and $J=1$.}\label{energy_p}
 \end{figure}
In  Fig. \ref{energy_p}(c), we plot the probability of the atom $\psi_e$, the left edge state $\psi_L$ and the right edge state $\psi_R$ as a function of $\theta$ for $|\phi_0\rangle$. We find that when $\theta=0.5\pi$, $G/G_{R,b}\gg 1$, $|\phi_0\rangle$ is mainly concentrated on the atomic excited state and With increasing $\theta$ up to $0.8\pi$, $G/G_{R,b}\to 0$ and  $|\phi_0\rangle$ is most concentrated on the left edge state as presented in the sub-Fig. \ref{energy_p}(c).  Thus by adiabatically tuning $\theta$, we can realize the transfer of atomic excitation to the left edge state. 
For $|\phi_\pm\rangle$, it is mainly concentrated in the excited of the atom and right-edge states when $\theta\to \pi$ as shown in  Fig. \ref{energy_p}(d) and we don't show it here.

Similarly when the atom is coupled to the sub-lattice $A$, there is still an eigenstate with zero eigenvalue, $|\phi_0\rangle=\{-G/G_{R,b},0,1\}$, which is a superposition of atom excited state and the right edge state.
By adjusting $\theta$, we can find that this as $\theta$ increases, the excitation is distributed mainly in the atomic excited state initially and will transfers to the right edge state finally. In addition, it is worth noting that when $\theta=\pi$, the left (right) edge states are mainly distributed on the leftmost (rightmost) sub-lattice $A$ ($B$), which implies that atoms in the excited state can be transferred to the end of the chain through SSH mode channels, and the direction is completely controllable by setting the atom coupling to sub-lattice A or B.
\begin{figure}[t]
  \centering
  \includegraphics[width=9cm]{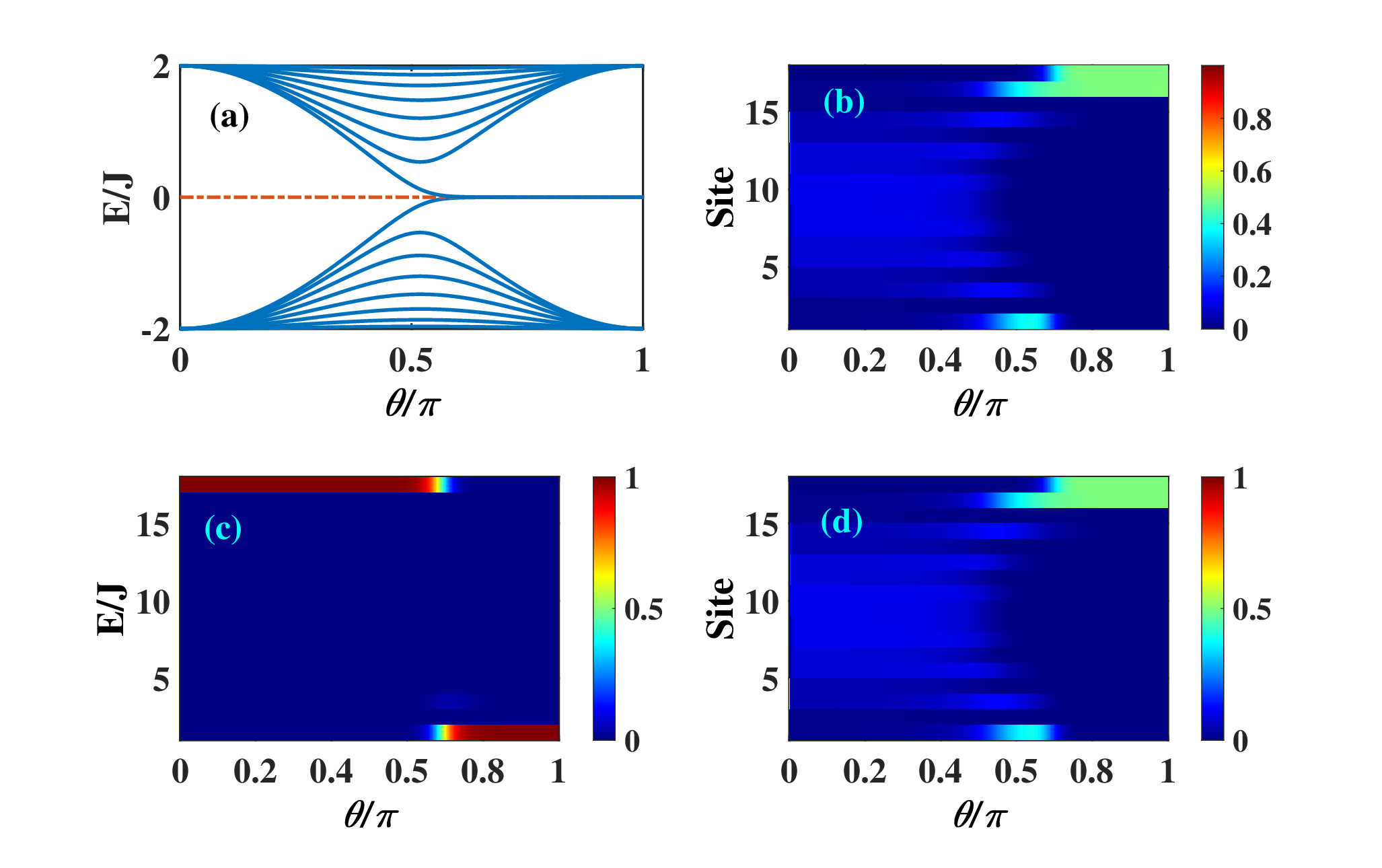}
  \caption{(a) The energy spectrum  with the variation of $\theta$ when the small atom couple with sub-lattice $B$. (b)(c)(d) The probability  distribution of the bandgap states marked in orange on the sites varies with $\theta$. The  parameters are setting to $N=8$, $g=0.01J$, $p=6$ and $J=1$.}\label{small_atom}
 \end{figure}

In order to validate our conclusions, in Figs. \ref{small_atom} (a), we plot the energy spectrum varying with $\theta$ under the case where the small atom couple to sub-lattice $B$ for the total Hamiltonian (\ref{ee1}).
It can be seen that there is three bandgap states, where  a zero energy state $|E_{N+1}\rangle$ in all ranges,  $|E_{N}\rangle$ and  $|E_{N+1}\rangle$ in  $\theta \in [0.5\pi, \pi]$.
In order to investigate the zero energy state $|E_{N+1}\rangle$, we plot the probability distribution of $|E_{N+1}\rangle$ on the sites varying with $\theta$ shown in  Fig. \ref{small_atom} (c). We find that in the range $\theta \in [0, 0.5\pi]$, the probability distribution of $|E_{N+1}\rangle$ is concentrated on the atom, and gradually transfer to the rightmost sub-lattice $A$. We have also investigated the distribution of the probability of the the other  bandgap states as shown in Figs. \ref{small_atom} (b)(d), it can be found that the distribution of the bandgap states is mainly concentrated on the excited of the atom and right-edge states when the SSH model in the topological phases, which is consistent with $|\phi_{\pm}\rangle$. That indicates that our previous discussion in the subspace is reasonable.

The above results mean that we can achieve controlled quantum information transfer. The  atom acts as the transmitter of the signal and the ends of the chain ($A_1, B_{2N}$) act as the receiver of the information.
The transfer of information to the leftmost and rightmost ends of the chain is achieved by controlling the atom coupling to sub-lattices $A$ or $B$, respectively. To further verify the feasibility of state transfer, we use the time-dependent Hamiltonian to evolve the initial state with $i\frac{d|\Psi\rangle}{dt}=H(\theta_t)|\Psi\rangle$, where we set $\theta_t=\Omega t$ with $\Omega$ is the the varying rate and that initially the atom is in an excited state and the topological chain is in the vacuum states as $\psi_i=\vert 0,0......0,1\rangle$.
\begin{figure}[h]
  \centering
  \includegraphics[width=9cm]{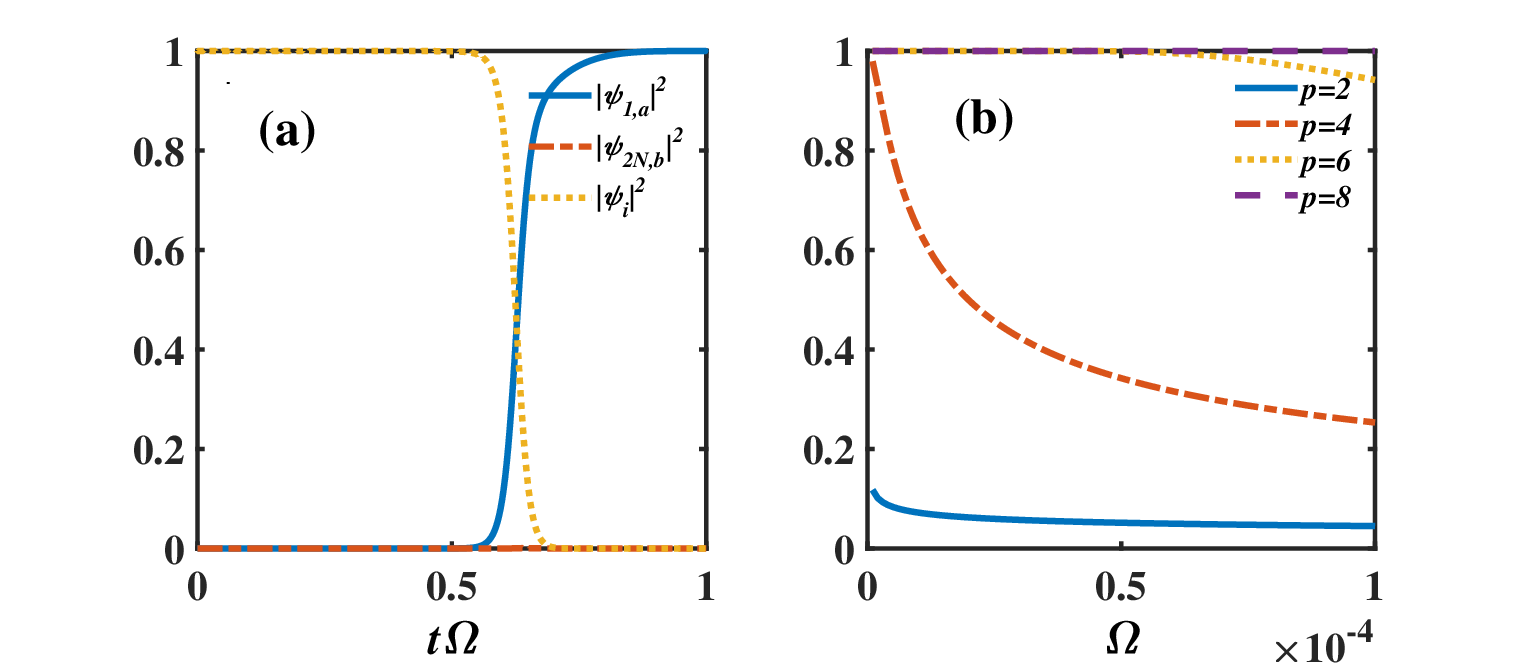}
  \caption{(a) The probability evolution with time for the  atom coupling with sub-lattice $B$ with $p=6$ and  \textcolor{blue}{$\Omega=10^{-5}$}. (b) The fidelity of the state transfer between the target state $\vert 1,0......0,0\rangle$ and the final state for different coupling position $p=2,4,6,8$.
  The  parameters are setting to $N=8$, $g=0.01J$ and $J=1$.}\label{evolution_ab}
 \end{figure}

The results of the numerical simulations are displayed in Fig. \ref{evolution_ab} (a), where $|\psi_{1,a}|^2$ and $|\psi_{2N,b}|^2$ the probability of  the leftmost and rightmost, respectively.
It can be seen that when the atom couple to sub-lattice $B$, the excitation is transferred to the rightmost sub-lattice $A$, while when atom couples to sub-lattice $A$, the excitation is transferred to the leftmost sub-lattice $B$.
However, during the evolution, the parameters $\theta$ need to be adiabatically varied to ensure a sufficiently high probability of success \cite{PhysRevB.103.085129,PhysRevA.102.022404}.  In Fig. \ref{evolution_ab} (b), we plot the fidelity of the state transfer between the target state $\vert 1,0......0,0\rangle$, where the leftmost sub-lattice $A$ is in the single excited state, and the final state for different coupling position. The numerical results show that as the atom approaches closer to the leftmost sub-lattice $B$, it does not require a sufficiently small coupling rate $\Omega$, however, when the atom is further away from the  leftmost  sub-lattice $B$, it requires enough  small coupling rate to achieve a high fidelity. 
This can be explained in terms of the three-state system, where the larger the coupling position $p$, the larger $G_{R,b}$, and the wider the width of the bandgap $\Delta_{\lambda}=|G_{R,b}|$ during the adiabatic process, so that a particularly small rate of change is not required, as long as it satisfies $g\Omega<\Delta^2_{\lambda}$ \cite{PhysRevA.98.012331}.

\section{excitation transfer with the adiabatic process  and entanglement dynamics}
In this section, we discuss the controllable excitation transfer beyond the adiabatic process. We use the atom coupling sub-lattice $B$ as an example. In the topological phase, when the atom in the excited state initially, under the Hamiltonian (\ref{H_{eff}}), the state will evolve as
\begin{equation}
  \begin{aligned}
   |\psi(t)\rangle=\alpha_e(t)|vac, e\rangle+\alpha_L(t)| \psi_{L}, g\rangle+\alpha_R(t)| \psi_{R}, g\rangle,
\end{aligned}
\end{equation}
with $\alpha_e (0) =1 $ and
\begin{equation}
\begin{aligned}
 \alpha_e(t)&= \frac{G^2+G_{R,b}^2\cos[\sqrt{G^2+G_{R,b}^2}t]}{G^2+G_{R,b}^2},\\
 \alpha_L(t)&=\frac{-GG_{R,b}+GG_{R,b}\cos[\sqrt{G^2+G_{R,b}^2}t])}{G^2+G_{R,b}^2},\\
 \alpha_R(t)&=\frac{iG_{R,b}\sin[\sqrt{G^2+G_{R,b}^2}t]}{\sqrt{G^2+G_{R,b}^2}}.\\
\end{aligned}
\end{equation}
\begin{figure}[t]
\centering
\includegraphics[width=9cm]{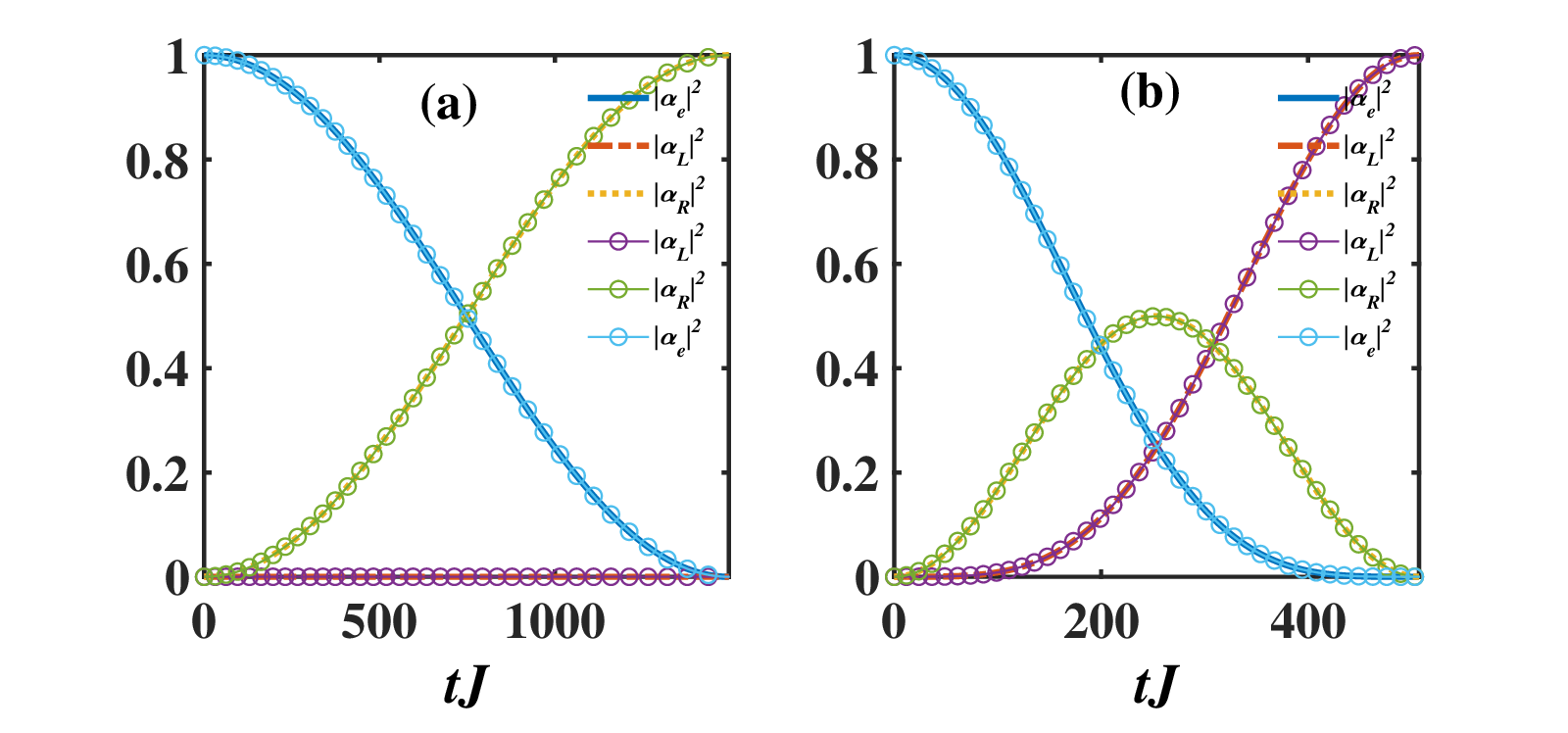}
\caption{(a)(b)The probability of the excited atom $|\alpha_e(t)|^2$ and the right and left edge states $|\alpha_{R,L}(t)|^2$  varying with time for $\theta=0.8\pi$ and   $\theta=0.6057\pi$, respectively.
The circle representation results from solving the full Hamiltonian dynamics evolution. The other parameters are $g=0.01J$, $p=7$ $N=8$ and $J=1$.}\label{ental}
 \end{figure}

Apparently, the above probability amplitudes depend on the $G$ and $G_{R,b}$.
As shown in Fig. \ref{energy_p} (a), when $\theta \approx  0.73\pi$, $G \approx 0$, thus the probability amplitude of left edge state $\alpha_L(t)=0$, the atom and the right edge state exhibit Rabi oscillations with the period depending on $G_{R,b}$. The result is displayed in  Fig. \ref{ental} (a).  For $0.5\pi<\theta<0.7\pi$ and  $p=7$, we find when $\theta=0.6057\pi$, $G=G_{R,b}$ the amplitudes becomes $\alpha_e(t)=(1+\cos(\sqrt{2}Gt))/2$, $\alpha_L(t)=(-1+\cos(\sqrt{2}Gt))/2$ and $\alpha_R(t)=i\sin(\sqrt{2}Gt)/\sqrt{2}$. It is evident that as t varies from 0 to $\pi/(\sqrt{2}G)$, the excitation is distinctly transferred from the atom to the left-edge state, as illustrated in Fig. \ref{ental}(b). 
In addition, when $t_f=\pi/\sqrt{G^2+G_{R,b}^2}$,  $\alpha_R(t_f)=0$, in this time $|\psi(t_f)\rangle$ is a superposition of the excited atomic state and the left edge state as $|\psi(t)\rangle=(\frac{G^2-G_{R,b}^2}{G^2+G_{R,b}^2}|10\rangle+ \frac{-2GG_{R,b}}{G^2+G_{R,b}^2}|01\rangle)|0_R\rangle$. Thus, by setting the coupling point and $\theta$, we can achieve the swap and entanglement between atom and left or right edge state.

\section{The effect of the disorder on the excitation transfer}

\begin{figure}[t]
  \centering
  \includegraphics[width=9cm]{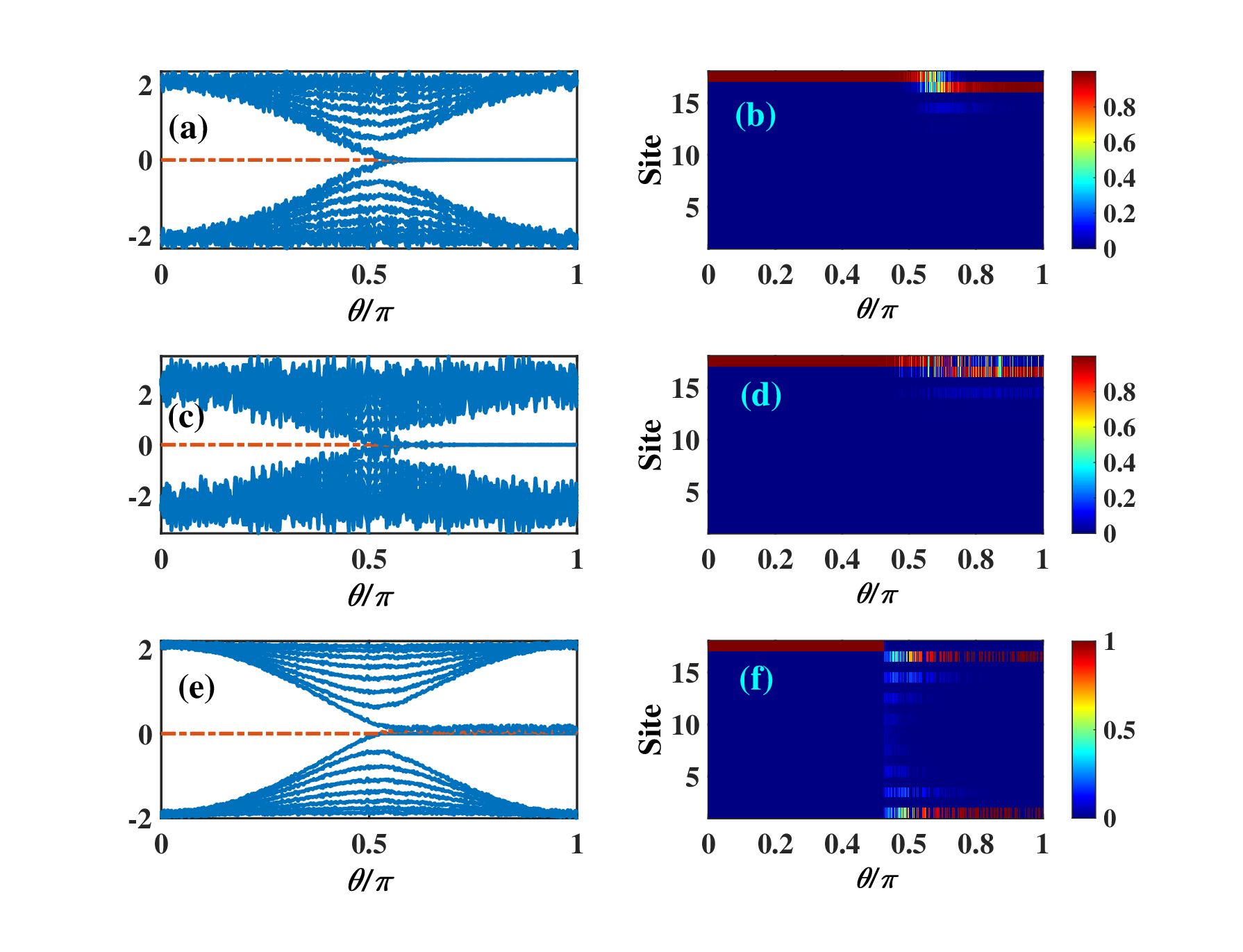}
  \caption{ The energy spectrum and  the probability of distribution of the bandgap state varying with $\theta$ under the diagonal and off-diagonal disorder.
  (a)(b))$\eta_{1,2}=0.2J$, $\varepsilon_{1,2}=0$.   (c)(d) $\eta_{1,2}=0.8J$, $\varepsilon_{1,2}=0$. (e)(f) $\eta_{1,2}=0$, $\varepsilon_{1,2}=0.2J$. The parameters are $g=0.01J$, $N=8$, $p=6$ and $J=1$.}\label{dis}
 \end{figure}

Next, we discuss the effect of disorder on the energy spectrum and  the probability of distribution of the bandgap state, where we mainly consider the effect of diagonal and off-diagonal disorder.
Under the consideration of disorder perturbations, the Hamiltonian  (\ref{ee1}) becomes
 \begin{equation}
  \begin{aligned}
   H&=\Delta \sigma^\dag\sigma+ \sum\limits_{i=1}^N \varepsilon_1  a_{i}^{\dagger}a_i+\varepsilon_2  b_{i}^{\dagger}b_i+(ga_{q}^{\dagger}\sigma+h.c.)
   \\&\quad+\sum\limits_{i=1}^N[(J_1+\eta_1) a_{i}^{\dagger}b_i + (J_2+\eta_2)a_{i+1}^{\dagger}b_i+h.c.],
  \end{aligned}
  \end{equation}
where  $\varepsilon_{1,2}$ and  $\eta_{1,2}$  are the diagonal and off-diagonal disorder. We consider $\varepsilon_{1,2}$ and  $\eta_{1,2}$  from random distributions with the range $[-\xi,\xi ]$, where $\xi$ is the disorder str. In Fig. (\ref{dis}),  we plot the energy spectrum and the probability of distribution of the bandgap state varying with $\theta$ under the diagonal and off-diagonal disorder. We can see that the presence of disorder affects both bands of the energy spectra, but the energy of the band gap states is not affected. For the weak non-diagonal perturbations, the probability distribution of the bandgap state is not influenced, but will be affected obviously by the perturbation strength a strong enough disorder comparing Figs. \ref{dis}(b)(d).  However, we find that the energy spectrum and the probability distribution of the band gap states are very sensitive to diagonal disorder as shown in Figs. \ref{dis}(e)(f).  Thus, for the present model, it is immune to weak off-diagonal disorder but extremely sensitive to off-diagonal disorder.

\section{discussion}
In this section, we discuss the experimental feasibility. In recent years, based on the flexibility of the parameters and the design capability of the system structure \cite{RevModPhys.93.025005,PhysRevLett.94.090501,doi:10.1126/science.1069372}, superconducting circuits have obtained much attention and have evolved into a well-established platform for the study of quantum simulation  \cite{doi:10.1126/science.1177838,PhysRevX.7.011016,Noh2017}, quantum computation \cite{Daley2022}, and quantum information processing \cite{PhysRevA.75.032329,Mirhosseini2019}.
 Our scheme can be implemented in the superconducting circuit systems. Recently, these works \cite{PhysRevX.11.011015,Bello2019,PhysRevResearch.2.012076,PhysRevX.9.011021,PhysRevLett.126.063601,Scigliuzzo2022} have pointed out theoretically and experimentally that each cell of SSH model can be mapped to the two  $LC$ resonators and the coupling of resonator sites is through an auxiliary capacitance and inductance in superconducting circuits, and the relative magnitude of intra- and inter-cell coupling between neighboring resonators is determined by the auxiliary capacitance and inductance  \cite{Houck2012,Schmidt2013}. By designing the intra- and extra-cell capacitance and inductance, a periodic modulation of the coupling can be achieved to obtain the SSH model.  Experimentally based on superconducting circuits, the coupling of atom to waveguides has been realized \cite{PhysRevA.95.053821,Kannan2020}.  Therefore, our scheme can be implemented with superconducting circuits in experiment.

\section{ conclusion}
 In this paper, we study the atom coupling to the finite size Su-Schrieffer-Heeger (SSH) model. We find that it can be regarded as the atom coupled to the waveguide with finite bandwidths and nonlinear dispersion relation when the finite SSH model is in the trivial phase. However, for the SSH model in the topological phase, when we consider the frequency of the atom is resonant with the SSH model, we find that the atom couples to edge states. In this case, we find that the special channels can realize the transfer of excited atoms to both ends of the chain by the adiabatic process. When the atom couples to the sub-lattice $B$ ($A$), the excitation of the atom can be transferred to the leftmost (rightmost) end of the chain. In addition, we can achieve excitation transfer from the excited state of the atom to the ends of the chain that without an adiabatic process. Our work provides a pathway for realizing controlled quantum information transfer based on the atom coupled topological matter.
\section{acknowledgments}

This work is supported by  National Natural Science Foundation of China (Grants No.12274053) and National Key R$\&$D Program of China (No.2021YFE0193500).

\bibliography{fouring}

\begin{thebibliography}{74}%
\makeatletter
\providecommand \@ifxundefined [1]{%
 \@ifx{#1\undefined}
}%
\providecommand \@ifnum [1]{%
 \ifnum #1\expandafter \@firstoftwo
 \else \expandafter \@secondoftwo
 \fi
}%
\providecommand \@ifx [1]{%
 \ifx #1\expandafter \@firstoftwo
 \else \expandafter \@secondoftwo
 \fi
}%
\providecommand \natexlab [1]{#1}%
\providecommand \enquote  [1]{``#1''}%
\providecommand \bibnamefont  [1]{#1}%
\providecommand \bibfnamefont [1]{#1}%
\providecommand \citenamefont [1]{#1}%
\providecommand \href@noop [0]{\@secondoftwo}%
\providecommand \href [0]{\begingroup \@sanitize@url \@href}%
\providecommand \@href[1]{\@@startlink{#1}\@@href}%
\providecommand \@@href[1]{\endgroup#1\@@endlink}%
\providecommand \@sanitize@url [0]{\catcode `\\12\catcode `\$12\catcode
  `\&12\catcode `\#12\catcode `\^12\catcode `\_12\catcode `\%12\relax}%
\providecommand \@@startlink[1]{}%
\providecommand \@@endlink[0]{}%
\providecommand \url  [0]{\begingroup\@sanitize@url \@url }%
\providecommand \@url [1]{\endgroup\@href {#1}{\urlprefix }}%
\providecommand \urlprefix  [0]{URL }%
\providecommand \Eprint [0]{\href }%
\providecommand \doibase [0]{https://doi.org/}%
\providecommand \selectlanguage [0]{\@gobble}%
\providecommand \bibinfo  [0]{\@secondoftwo}%
\providecommand \bibfield  [0]{\@secondoftwo}%
\providecommand \translation [1]{[#1]}%
\providecommand \BibitemOpen [0]{}%
\providecommand \bibitemStop [0]{}%
\providecommand \bibitemNoStop [0]{.\EOS\space}%
\providecommand \EOS [0]{\spacefactor3000\relax}%
\providecommand \BibitemShut  [1]{\csname bibitem#1\endcsname}%
\let\auto@bib@innerbib\@empty
\bibitem [{\citenamefont {Hasan}\ and\ \citenamefont
  {Kane}(2010)}]{RevModPhys.82.3045}%
  \BibitemOpen
  \bibfield  {author} {\bibinfo {author} {\bibfnamefont {M.~Z.}\ \bibnamefont
  {Hasan}}\ and\ \bibinfo {author} {\bibfnamefont {C.~L.}\ \bibnamefont
  {Kane}},\ }\bibfield  {title} {\bibinfo {title} {Colloquium: Topological
  insulators},\ }\href {https://doi.org/10.1103/RevModPhys.82.3045} {\bibfield
  {journal} {\bibinfo  {journal} {Rev. Mod. Phys.}\ }\textbf {\bibinfo {volume}
  {82}},\ \bibinfo {pages} {3045} (\bibinfo {year} {2010})}\BibitemShut
  {NoStop}%
\bibitem [{\citenamefont {Wray}\ \emph {et~al.}(2010)\citenamefont {Wray},
  \citenamefont {Xu}, \citenamefont {Xia}, \citenamefont {Hor}, \citenamefont
  {Qian}, \citenamefont {Fedorov}, \citenamefont {Lin}, \citenamefont {Bansil},
  \citenamefont {Cava},\ and\ \citenamefont {Hasan}}]{Wray2010}%
  \BibitemOpen
  \bibfield  {author} {\bibinfo {author} {\bibfnamefont {L.~A.}\ \bibnamefont
  {Wray}}, \bibinfo {author} {\bibfnamefont {S.~Y.}\ \bibnamefont {Xu}},
  \bibinfo {author} {\bibfnamefont {Y.}~\bibnamefont {Xia}}, \bibinfo {author}
  {\bibfnamefont {Y.~S.}\ \bibnamefont {Hor}}, \bibinfo {author} {\bibfnamefont
  {D.}~\bibnamefont {Qian}}, \bibinfo {author} {\bibfnamefont {A.~V.}\
  \bibnamefont {Fedorov}}, \bibinfo {author} {\bibfnamefont {H.}~\bibnamefont
  {Lin}}, \bibinfo {author} {\bibfnamefont {A.}~\bibnamefont {Bansil}},
  \bibinfo {author} {\bibfnamefont {R.~J.}\ \bibnamefont {Cava}},\ and\
  \bibinfo {author} {\bibfnamefont {M.~Z.}\ \bibnamefont {Hasan}},\ }\bibfield
  {title} {\bibinfo {title} {{Observation of topological order in a
  superconducting doped topological insulator}},\ }\href
  {https://doi.org/10.1038/nphys1762} {\bibfield  {journal} {\bibinfo
  {journal} {Nat. Phys.}\ }\textbf {\bibinfo {volume} {6}},\ \bibinfo {pages}
  {855} (\bibinfo {year} {2010})}\BibitemShut {NoStop}%
\bibitem [{\citenamefont {Shalaev}\ \emph {et~al.}(2019)\citenamefont
  {Shalaev}, \citenamefont {Walasik}, \citenamefont {Tsukernik}, \citenamefont
  {Xu},\ and\ \citenamefont {Litchinitser}}]{Shalaev2019}%
  \BibitemOpen
  \bibfield  {author} {\bibinfo {author} {\bibfnamefont {M.~I.}\ \bibnamefont
  {Shalaev}}, \bibinfo {author} {\bibfnamefont {W.}~\bibnamefont {Walasik}},
  \bibinfo {author} {\bibfnamefont {A.}~\bibnamefont {Tsukernik}}, \bibinfo
  {author} {\bibfnamefont {Y.}~\bibnamefont {Xu}},\ and\ \bibinfo {author}
  {\bibfnamefont {N.~M.}\ \bibnamefont {Litchinitser}},\ }\bibfield  {title}
  {\bibinfo {title} {{Robust topologically protected transport in photonic
  crystals at telecommunication wavelengths}},\ }\href
  {https://doi.org/10.1038/s41565-018-0297-6} {\bibfield  {journal} {\bibinfo
  {journal} {Nat. Nanotechnol.}\ }\textbf {\bibinfo {volume} {14}},\ \bibinfo
  {pages} {31} (\bibinfo {year} {2019})}\BibitemShut {NoStop}%
\bibitem [{\citenamefont {Ozawa}\ \emph {et~al.}(2019)\citenamefont {Ozawa},
  \citenamefont {Price}, \citenamefont {Amo}, \citenamefont {Goldman},
  \citenamefont {Hafezi}, \citenamefont {Lu}, \citenamefont {Rechtsman},
  \citenamefont {Schuster}, \citenamefont {Simon}, \citenamefont {Zilberberg},\
  and\ \citenamefont {Carusotto}}]{RevModPhys.91.015006}%
  \BibitemOpen
  \bibfield  {author} {\bibinfo {author} {\bibfnamefont {T.}~\bibnamefont
  {Ozawa}}, \bibinfo {author} {\bibfnamefont {H.~M.}\ \bibnamefont {Price}},
  \bibinfo {author} {\bibfnamefont {A.}~\bibnamefont {Amo}}, \bibinfo {author}
  {\bibfnamefont {N.}~\bibnamefont {Goldman}}, \bibinfo {author} {\bibfnamefont
  {M.}~\bibnamefont {Hafezi}}, \bibinfo {author} {\bibfnamefont
  {L.}~\bibnamefont {Lu}}, \bibinfo {author} {\bibfnamefont {M.~C.}\
  \bibnamefont {Rechtsman}}, \bibinfo {author} {\bibfnamefont {D.}~\bibnamefont
  {Schuster}}, \bibinfo {author} {\bibfnamefont {J.}~\bibnamefont {Simon}},
  \bibinfo {author} {\bibfnamefont {O.}~\bibnamefont {Zilberberg}},\ and\
  \bibinfo {author} {\bibfnamefont {I.}~\bibnamefont {Carusotto}},\ }\bibfield
  {title} {\bibinfo {title} {Topological photonics},\ }\href
  {https://doi.org/10.1103/RevModPhys.91.015006} {\bibfield  {journal}
  {\bibinfo  {journal} {Rev. Mod. Phys.}\ }\textbf {\bibinfo {volume} {91}},\
  \bibinfo {pages} {015006} (\bibinfo {year} {2019})}\BibitemShut {NoStop}%
\bibitem [{\citenamefont {Goblot}\ \emph {et~al.}(2017)\citenamefont {Goblot},
  \citenamefont {Galopin}, \citenamefont {Lema{\^{i}}tre}, \citenamefont
  {Ozawa}, \citenamefont {Gratiet}, \citenamefont {Sagnes}, \citenamefont
  {Bloch},\ and\ \citenamefont {Amo}}]{Goblot2017}%
  \BibitemOpen
  \bibfield  {author} {\bibinfo {author} {\bibfnamefont {V.}~\bibnamefont
  {Goblot}}, \bibinfo {author} {\bibfnamefont {E.}~\bibnamefont {Galopin}},
  \bibinfo {author} {\bibfnamefont {A.}~\bibnamefont {Lema{\^{i}}tre}},
  \bibinfo {author} {\bibfnamefont {T.}~\bibnamefont {Ozawa}}, \bibinfo
  {author} {\bibfnamefont {L.~L.}\ \bibnamefont {Gratiet}}, \bibinfo {author}
  {\bibfnamefont {I.}~\bibnamefont {Sagnes}}, \bibinfo {author} {\bibfnamefont
  {J.}~\bibnamefont {Bloch}},\ and\ \bibinfo {author} {\bibfnamefont
  {A.}~\bibnamefont {Amo}},\ }\bibfield  {title} {\bibinfo {title} {{Lasing in
  topological edge states of a one-dimensional lattice}},\ }\href
  {https://doi.org/10.1038/s41566-017-0006-2} {\bibfield  {journal} {\bibinfo
  {journal} {Nat. Photonics}\ }\textbf {\bibinfo {volume} {11}},\ \bibinfo
  {pages} {1} (\bibinfo {year} {2017})}\BibitemShut {NoStop}%
\bibitem [{\citenamefont {Parto}\ \emph {et~al.}(2018)\citenamefont {Parto},
  \citenamefont {Wittek}, \citenamefont {Hodaei}, \citenamefont {Harari},
  \citenamefont {Bandres}, \citenamefont {Ren}, \citenamefont {Rechtsman},
  \citenamefont {Segev}, \citenamefont {Christodoulides},\ and\ \citenamefont
  {Khajavikhan}}]{PhysRevLett.120.113901}%
  \BibitemOpen
  \bibfield  {author} {\bibinfo {author} {\bibfnamefont {M.}~\bibnamefont
  {Parto}}, \bibinfo {author} {\bibfnamefont {S.}~\bibnamefont {Wittek}},
  \bibinfo {author} {\bibfnamefont {H.}~\bibnamefont {Hodaei}}, \bibinfo
  {author} {\bibfnamefont {G.}~\bibnamefont {Harari}}, \bibinfo {author}
  {\bibfnamefont {M.~A.}\ \bibnamefont {Bandres}}, \bibinfo {author}
  {\bibfnamefont {J.}~\bibnamefont {Ren}}, \bibinfo {author} {\bibfnamefont
  {M.~C.}\ \bibnamefont {Rechtsman}}, \bibinfo {author} {\bibfnamefont
  {M.}~\bibnamefont {Segev}}, \bibinfo {author} {\bibfnamefont {D.~N.}\
  \bibnamefont {Christodoulides}},\ and\ \bibinfo {author} {\bibfnamefont
  {M.}~\bibnamefont {Khajavikhan}},\ }\bibfield  {title} {\bibinfo {title}
  {Edge-mode lasing in 1d topological active arrays},\ }\href
  {https://doi.org/10.1103/PhysRevLett.120.113901} {\bibfield  {journal}
  {\bibinfo  {journal} {Phys. Rev. Lett.}\ }\textbf {\bibinfo {volume} {120}},\
  \bibinfo {pages} {113901} (\bibinfo {year} {2018})}\BibitemShut {NoStop}%
\bibitem [{\citenamefont {Zhang}\ \emph {et~al.}(2022)\citenamefont {Zhang},
  \citenamefont {Liu}, \citenamefont {Zhang}, \citenamefont {Gong},
  \citenamefont {Zhang}, \citenamefont {Yan}, \citenamefont {Su}, \citenamefont
  {Jing},\ and\ \citenamefont {Feng}}]{Zhang2022}%
  \BibitemOpen
  \bibfield  {author} {\bibinfo {author} {\bibfnamefont {J.~Q.}\ \bibnamefont
  {Zhang}}, \bibinfo {author} {\bibfnamefont {J.~X.}\ \bibnamefont {Liu}},
  \bibinfo {author} {\bibfnamefont {H.~L.}\ \bibnamefont {Zhang}}, \bibinfo
  {author} {\bibfnamefont {Z.~R.}\ \bibnamefont {Gong}}, \bibinfo {author}
  {\bibfnamefont {S.}~\bibnamefont {Zhang}}, \bibinfo {author} {\bibfnamefont
  {L.~L.}\ \bibnamefont {Yan}}, \bibinfo {author} {\bibfnamefont {S.~L.}\
  \bibnamefont {Su}}, \bibinfo {author} {\bibfnamefont {H.}~\bibnamefont
  {Jing}},\ and\ \bibinfo {author} {\bibfnamefont {M.}~\bibnamefont {Feng}},\
  }\bibfield  {title} {\bibinfo {title} {{Topological optomechanical amplifier
  in synthetic PT -symmetry}},\ }\href
  {https://doi.org/10.1515/nanoph-2021-0721} {\bibfield  {journal} {\bibinfo
  {journal} {Nanophotonics}\ }\textbf {\bibinfo {volume} {11}},\ \bibinfo
  {pages} {1149} (\bibinfo {year} {2022})}\BibitemShut {NoStop}%
\bibitem [{\citenamefont {Ramos}\ \emph {et~al.}(2021)\citenamefont {Ramos},
  \citenamefont {Garc\'{\i}a-Ripoll},\ and\ \citenamefont
  {Porras}}]{PhysRevA.103.033513}%
  \BibitemOpen
  \bibfield  {author} {\bibinfo {author} {\bibfnamefont {T.}~\bibnamefont
  {Ramos}}, \bibinfo {author} {\bibfnamefont {J.~J.}\ \bibnamefont
  {Garc\'{\i}a-Ripoll}},\ and\ \bibinfo {author} {\bibfnamefont
  {D.}~\bibnamefont {Porras}},\ }\bibfield  {title} {\bibinfo {title}
  {Topological input-output theory for directional amplification},\ }\href
  {https://doi.org/10.1103/PhysRevA.103.033513} {\bibfield  {journal} {\bibinfo
   {journal} {Phys. Rev. A}\ }\textbf {\bibinfo {volume} {103}},\ \bibinfo
  {pages} {033513} (\bibinfo {year} {2021})}\BibitemShut {NoStop}%
\bibitem [{\citenamefont {Wanjura}\ \emph {et~al.}(2020)\citenamefont
  {Wanjura}, \citenamefont {Brunelli},\ and\ \citenamefont
  {Nunnenkamp}}]{Wanjura2020}%
  \BibitemOpen
  \bibfield  {author} {\bibinfo {author} {\bibfnamefont {C.~C.}\ \bibnamefont
  {Wanjura}}, \bibinfo {author} {\bibfnamefont {M.}~\bibnamefont {Brunelli}},\
  and\ \bibinfo {author} {\bibfnamefont {A.}~\bibnamefont {Nunnenkamp}},\
  }\bibfield  {title} {\bibinfo {title} {{Topological framework for directional
  amplification in driven-dissipative cavity arrays}},\ }\href
  {https://doi.org/10.1038/s41467-020-16863-9} {\bibfield  {journal} {\bibinfo
  {journal} {Nat. Commun.}\ }\textbf {\bibinfo {volume} {11}},\ \bibinfo
  {pages} {1} (\bibinfo {year} {2020})}\BibitemShut {NoStop}%
\bibitem [{\citenamefont {Porras}\ and\ \citenamefont
  {Fern\'andez-Lorenzo}(2019)}]{PhysRevLett.122.143901}%
  \BibitemOpen
  \bibfield  {author} {\bibinfo {author} {\bibfnamefont {D.}~\bibnamefont
  {Porras}}\ and\ \bibinfo {author} {\bibfnamefont {S.}~\bibnamefont
  {Fern\'andez-Lorenzo}},\ }\bibfield  {title} {\bibinfo {title} {Topological
  amplification in photonic lattices},\ }\href
  {https://doi.org/10.1103/PhysRevLett.122.143901} {\bibfield  {journal}
  {\bibinfo  {journal} {Phys. Rev. Lett.}\ }\textbf {\bibinfo {volume} {122}},\
  \bibinfo {pages} {143901} (\bibinfo {year} {2019})}\BibitemShut {NoStop}%
\bibitem [{\citenamefont {Su}\ \emph {et~al.}(1979)\citenamefont {Su},
  \citenamefont {Schrieffer},\ and\ \citenamefont
  {Heeger}}]{PhysRevLett.42.1698}%
  \BibitemOpen
  \bibfield  {author} {\bibinfo {author} {\bibfnamefont {W.~P.}\ \bibnamefont
  {Su}}, \bibinfo {author} {\bibfnamefont {J.~R.}\ \bibnamefont {Schrieffer}},\
  and\ \bibinfo {author} {\bibfnamefont {A.~J.}\ \bibnamefont {Heeger}},\
  }\bibfield  {title} {\bibinfo {title} {Solitons in polyacetylene},\ }\href
  {https://doi.org/10.1103/PhysRevLett.42.1698} {\bibfield  {journal} {\bibinfo
   {journal} {Phys. Rev. Lett.}\ }\textbf {\bibinfo {volume} {42}},\ \bibinfo
  {pages} {1698} (\bibinfo {year} {1979})}\BibitemShut {NoStop}%
\bibitem [{\citenamefont {Heeger}\ \emph {et~al.}(1988)\citenamefont {Heeger},
  \citenamefont {Kivelson}, \citenamefont {Schrieffer},\ and\ \citenamefont
  {Su}}]{RevModPhys.60.781}%
  \BibitemOpen
  \bibfield  {author} {\bibinfo {author} {\bibfnamefont {A.~J.}\ \bibnamefont
  {Heeger}}, \bibinfo {author} {\bibfnamefont {S.}~\bibnamefont {Kivelson}},
  \bibinfo {author} {\bibfnamefont {J.~R.}\ \bibnamefont {Schrieffer}},\ and\
  \bibinfo {author} {\bibfnamefont {W.~P.}\ \bibnamefont {Su}},\ }\bibfield
  {title} {\bibinfo {title} {Solitons in conducting polymers},\ }\href
  {https://doi.org/10.1103/RevModPhys.60.781} {\bibfield  {journal} {\bibinfo
  {journal} {Rev. Mod. Phys.}\ }\textbf {\bibinfo {volume} {60}},\ \bibinfo
  {pages} {781} (\bibinfo {year} {1988})}\BibitemShut {NoStop}%
\bibitem [{\citenamefont {Bansil}\ \emph {et~al.}(2016)\citenamefont {Bansil},
  \citenamefont {Lin},\ and\ \citenamefont {Das}}]{RevModPhys.88.021004}%
  \BibitemOpen
  \bibfield  {author} {\bibinfo {author} {\bibfnamefont {A.}~\bibnamefont
  {Bansil}}, \bibinfo {author} {\bibfnamefont {H.}~\bibnamefont {Lin}},\ and\
  \bibinfo {author} {\bibfnamefont {T.}~\bibnamefont {Das}},\ }\bibfield
  {title} {\bibinfo {title} {Colloquium: Topological band theory},\ }\href
  {https://doi.org/10.1103/RevModPhys.88.021004} {\bibfield  {journal}
  {\bibinfo  {journal} {Rev. Mod. Phys.}\ }\textbf {\bibinfo {volume} {88}},\
  \bibinfo {pages} {021004} (\bibinfo {year} {2016})}\BibitemShut {NoStop}%
\bibitem [{\citenamefont {Xu}\ \emph {et~al.}(2022)\citenamefont {Xu},
  \citenamefont {Zhao}, \citenamefont {Wang}, \citenamefont {Chen},\ and\
  \citenamefont {Liu}}]{10.3389/fphy.2021.813801}%
  \BibitemOpen
  \bibfield  {author} {\bibinfo {author} {\bibfnamefont {X.-W.}\ \bibnamefont
  {Xu}}, \bibinfo {author} {\bibfnamefont {Y.-J.}\ \bibnamefont {Zhao}},
  \bibinfo {author} {\bibfnamefont {H.}~\bibnamefont {Wang}}, \bibinfo {author}
  {\bibfnamefont {A.-X.}\ \bibnamefont {Chen}},\ and\ \bibinfo {author}
  {\bibfnamefont {Y.-X.}\ \bibnamefont {Liu}},\ }\bibfield  {title} {\bibinfo
  {title} {Generalized su-schrieffer-heeger model in one dimensional
  optomechanical arrays},\ }\href
  {https://www.frontiersin.org/article/10.3389/fphy.2021.813801} {\bibfield
  {journal} {\bibinfo  {journal} {Front. Phys.}\ }\textbf {\bibinfo {volume}
  {9}} (\bibinfo {year} {2022})}\BibitemShut {NoStop}%
\bibitem [{\citenamefont {Meier}\ \emph {et~al.}(2016)\citenamefont {Meier},
  \citenamefont {An},\ and\ \citenamefont {Gadway}}]{Meier2016}%
  \BibitemOpen
  \bibfield  {author} {\bibinfo {author} {\bibfnamefont {E.~J.}\ \bibnamefont
  {Meier}}, \bibinfo {author} {\bibfnamefont {F.~A.}\ \bibnamefont {An}},\ and\
  \bibinfo {author} {\bibfnamefont {B.}~\bibnamefont {Gadway}},\ }\bibfield
  {title} {\bibinfo {title} {{Observation of the topological soliton state in
  the Su-Schrieffer-Heeger model}},\ }\href
  {https://doi.org/10.1038/ncomms13986} {\bibfield  {journal} {\bibinfo
  {journal} {Nat. Commun.}\ }\textbf {\bibinfo {volume} {7}},\ \bibinfo {pages}
  {13986} (\bibinfo {year} {2016})}\BibitemShut {NoStop}%
\bibitem [{\citenamefont {Tian}\ \emph {et~al.}(2022)\citenamefont {Tian},
  \citenamefont {Zhang}, \citenamefont {Zhang}, \citenamefont {Wu},
  \citenamefont {Lin}, \citenamefont {Zhou}, \citenamefont {Duan},
  \citenamefont {Jiang},\ and\ \citenamefont {Du}}]{PhysRevLett.129.215901}%
  \BibitemOpen
  \bibfield  {author} {\bibinfo {author} {\bibfnamefont {T.}~\bibnamefont
  {Tian}}, \bibinfo {author} {\bibfnamefont {Y.}~\bibnamefont {Zhang}},
  \bibinfo {author} {\bibfnamefont {L.}~\bibnamefont {Zhang}}, \bibinfo
  {author} {\bibfnamefont {L.}~\bibnamefont {Wu}}, \bibinfo {author}
  {\bibfnamefont {S.}~\bibnamefont {Lin}}, \bibinfo {author} {\bibfnamefont
  {J.}~\bibnamefont {Zhou}}, \bibinfo {author} {\bibfnamefont {C.-K.}\
  \bibnamefont {Duan}}, \bibinfo {author} {\bibfnamefont {J.-H.}\ \bibnamefont
  {Jiang}},\ and\ \bibinfo {author} {\bibfnamefont {J.}~\bibnamefont {Du}},\
  }\bibfield  {title} {\bibinfo {title} {Experimental realization of
  nonreciprocal adiabatic transfer of phonons in a dynamically modulated
  nanomechanical topological insulator},\ }\href
  {https://doi.org/10.1103/PhysRevLett.129.215901} {\bibfield  {journal}
  {\bibinfo  {journal} {Phys. Rev. Lett.}\ }\textbf {\bibinfo {volume} {129}},\
  \bibinfo {pages} {215901} (\bibinfo {year} {2022})}\BibitemShut {NoStop}%
\bibitem [{\citenamefont {Palaiodimopoulos}\ \emph {et~al.}(2021)\citenamefont
  {Palaiodimopoulos}, \citenamefont {Brouzos}, \citenamefont {Diakonos},\ and\
  \citenamefont {Theocharis}}]{PhysRevA.103.052409}%
  \BibitemOpen
  \bibfield  {author} {\bibinfo {author} {\bibfnamefont {N.~E.}\ \bibnamefont
  {Palaiodimopoulos}}, \bibinfo {author} {\bibfnamefont {I.}~\bibnamefont
  {Brouzos}}, \bibinfo {author} {\bibfnamefont {F.~K.}\ \bibnamefont
  {Diakonos}},\ and\ \bibinfo {author} {\bibfnamefont {G.}~\bibnamefont
  {Theocharis}},\ }\bibfield  {title} {\bibinfo {title} {Fast and robust
  quantum state transfer via a topological chain},\ }\href
  {https://doi.org/10.1103/PhysRevA.103.052409} {\bibfield  {journal} {\bibinfo
   {journal} {Phys. Rev. A}\ }\textbf {\bibinfo {volume} {103}},\ \bibinfo
  {pages} {052409} (\bibinfo {year} {2021})}\BibitemShut {NoStop}%
\bibitem [{\citenamefont {Mei}\ \emph {et~al.}(2018)\citenamefont {Mei},
  \citenamefont {Chen}, \citenamefont {Tian}, \citenamefont {Zhu},\ and\
  \citenamefont {Jia}}]{PhysRevA.98.012331}%
  \BibitemOpen
  \bibfield  {author} {\bibinfo {author} {\bibfnamefont {F.}~\bibnamefont
  {Mei}}, \bibinfo {author} {\bibfnamefont {G.}~\bibnamefont {Chen}}, \bibinfo
  {author} {\bibfnamefont {L.}~\bibnamefont {Tian}}, \bibinfo {author}
  {\bibfnamefont {S.-L.}\ \bibnamefont {Zhu}},\ and\ \bibinfo {author}
  {\bibfnamefont {S.}~\bibnamefont {Jia}},\ }\bibfield  {title} {\bibinfo
  {title} {Robust quantum state transfer via topological edge states in
  superconducting qubit chains},\ }\href
  {https://doi.org/10.1103/PhysRevA.98.012331} {\bibfield  {journal} {\bibinfo
  {journal} {Phys. Rev. A}\ }\textbf {\bibinfo {volume} {98}},\ \bibinfo
  {pages} {012331} (\bibinfo {year} {2018})}\BibitemShut {NoStop}%
\bibitem [{\citenamefont {Wang}\ \emph
  {et~al.}(2022{\natexlab{a}})\citenamefont {Wang}, \citenamefont {Li},
  \citenamefont {Gong},\ and\ \citenamefont {Liu}}]{PhysRevA.106.052411}%
  \BibitemOpen
  \bibfield  {author} {\bibinfo {author} {\bibfnamefont {C.}~\bibnamefont
  {Wang}}, \bibinfo {author} {\bibfnamefont {L.}~\bibnamefont {Li}}, \bibinfo
  {author} {\bibfnamefont {J.}~\bibnamefont {Gong}},\ and\ \bibinfo {author}
  {\bibfnamefont {Y.-x.}\ \bibnamefont {Liu}},\ }\bibfield  {title} {\bibinfo
  {title} {Arbitrary entangled state transfer via a topological qubit chain},\
  }\href {https://doi.org/10.1103/PhysRevA.106.052411} {\bibfield  {journal}
  {\bibinfo  {journal} {Phys. Rev. A}\ }\textbf {\bibinfo {volume} {106}},\
  \bibinfo {pages} {052411} (\bibinfo {year} {2022}{\natexlab{a}})}\BibitemShut
  {NoStop}%
\bibitem [{\citenamefont {Qi}\ \emph {et~al.}(2020)\citenamefont {Qi},
  \citenamefont {Wang}, \citenamefont {Liu}, \citenamefont {Zhang},\ and\
  \citenamefont {Wang}}]{PhysRevA.102.022404}%
  \BibitemOpen
  \bibfield  {author} {\bibinfo {author} {\bibfnamefont {L.}~\bibnamefont
  {Qi}}, \bibinfo {author} {\bibfnamefont {G.-L.}\ \bibnamefont {Wang}},
  \bibinfo {author} {\bibfnamefont {S.}~\bibnamefont {Liu}}, \bibinfo {author}
  {\bibfnamefont {S.}~\bibnamefont {Zhang}},\ and\ \bibinfo {author}
  {\bibfnamefont {H.-F.}\ \bibnamefont {Wang}},\ }\bibfield  {title} {\bibinfo
  {title} {Engineering the topological state transfer and topological beam
  splitter in an even-sized su-schrieffer-heeger chain},\ }\href
  {https://doi.org/10.1103/PhysRevA.102.022404} {\bibfield  {journal} {\bibinfo
   {journal} {Phys. Rev. A}\ }\textbf {\bibinfo {volume} {102}},\ \bibinfo
  {pages} {022404} (\bibinfo {year} {2020})}\BibitemShut {NoStop}%
\bibitem [{\citenamefont {Cao}\ \emph {et~al.}(2021)\citenamefont {Cao},
  \citenamefont {Cui}, \citenamefont {Yi},\ and\ \citenamefont
  {Wang}}]{PhysRevA.103.023504}%
  \BibitemOpen
  \bibfield  {author} {\bibinfo {author} {\bibfnamefont {J.}~\bibnamefont
  {Cao}}, \bibinfo {author} {\bibfnamefont {W.-X.}\ \bibnamefont {Cui}},
  \bibinfo {author} {\bibfnamefont {X.~X.}\ \bibnamefont {Yi}},\ and\ \bibinfo
  {author} {\bibfnamefont {H.-F.}\ \bibnamefont {Wang}},\ }\bibfield  {title}
  {\bibinfo {title} {Controllable photon-phonon conversion via the
  topologically protected edge channel in an optomechanical lattice},\ }\href
  {https://doi.org/10.1103/PhysRevA.103.023504} {\bibfield  {journal} {\bibinfo
   {journal} {Phys. Rev. A}\ }\textbf {\bibinfo {volume} {103}},\ \bibinfo
  {pages} {023504} (\bibinfo {year} {2021})}\BibitemShut {NoStop}%
\bibitem [{\citenamefont {Zheng}\ \emph {et~al.}(2022)\citenamefont {Zheng},
  \citenamefont {Yi},\ and\ \citenamefont {Wang}}]{PhysRevApplied.18.054037}%
  \BibitemOpen
  \bibfield  {author} {\bibinfo {author} {\bibfnamefont {L.-N.}\ \bibnamefont
  {Zheng}}, \bibinfo {author} {\bibfnamefont {X.}~\bibnamefont {Yi}},\ and\
  \bibinfo {author} {\bibfnamefont {H.-F.}\ \bibnamefont {Wang}},\ }\bibfield
  {title} {\bibinfo {title} {Engineering a phase-robust topological router in a
  dimerized superconducting-circuit lattice with long-range hopping and chiral
  symmetry},\ }\href {https://doi.org/10.1103/PhysRevApplied.18.054037}
  {\bibfield  {journal} {\bibinfo  {journal} {Phys. Rev. Appl.}\ }\textbf
  {\bibinfo {volume} {18}},\ \bibinfo {pages} {054037} (\bibinfo {year}
  {2022})}\BibitemShut {NoStop}%
\bibitem [{\citenamefont {Qi}\ \emph {et~al.}(2023)\citenamefont {Qi},
  \citenamefont {Han}, \citenamefont {Liu}, \citenamefont {Wang},\ and\
  \citenamefont {He}}]{PhysRevA.107.062214}%
  \BibitemOpen
  \bibfield  {author} {\bibinfo {author} {\bibfnamefont {L.}~\bibnamefont
  {Qi}}, \bibinfo {author} {\bibfnamefont {N.}~\bibnamefont {Han}}, \bibinfo
  {author} {\bibfnamefont {S.}~\bibnamefont {Liu}}, \bibinfo {author}
  {\bibfnamefont {H.-F.}\ \bibnamefont {Wang}},\ and\ \bibinfo {author}
  {\bibfnamefont {A.-L.}\ \bibnamefont {He}},\ }\bibfield  {title} {\bibinfo
  {title} {Controllable excitation transmission and topological switch based on
  an imaginary topological channel in a non-hermitian su-schrieffer-heeger
  chain},\ }\href {https://doi.org/10.1103/PhysRevA.107.062214} {\bibfield
  {journal} {\bibinfo  {journal} {Phys. Rev. A}\ }\textbf {\bibinfo {volume}
  {107}},\ \bibinfo {pages} {062214} (\bibinfo {year} {2023})}\BibitemShut
  {NoStop}%
\bibitem [{\citenamefont {Qi}\ \emph {et~al.}(2021{\natexlab{a}})\citenamefont
  {Qi}, \citenamefont {Yan}, \citenamefont {Xing}, \citenamefont {Zhao},
  \citenamefont {Liu}, \citenamefont {Cui}, \citenamefont {Han}, \citenamefont
  {Zhang},\ and\ \citenamefont {Wang}}]{PhysRevResearch.3.023037}%
  \BibitemOpen
  \bibfield  {author} {\bibinfo {author} {\bibfnamefont {L.}~\bibnamefont
  {Qi}}, \bibinfo {author} {\bibfnamefont {Y.}~\bibnamefont {Yan}}, \bibinfo
  {author} {\bibfnamefont {Y.}~\bibnamefont {Xing}}, \bibinfo {author}
  {\bibfnamefont {X.-D.}\ \bibnamefont {Zhao}}, \bibinfo {author}
  {\bibfnamefont {S.}~\bibnamefont {Liu}}, \bibinfo {author} {\bibfnamefont
  {W.-X.}\ \bibnamefont {Cui}}, \bibinfo {author} {\bibfnamefont
  {X.}~\bibnamefont {Han}}, \bibinfo {author} {\bibfnamefont {S.}~\bibnamefont
  {Zhang}},\ and\ \bibinfo {author} {\bibfnamefont {H.-F.}\ \bibnamefont
  {Wang}},\ }\bibfield  {title} {\bibinfo {title} {Topological router induced
  via long-range hopping in a su-schrieffer-heeger chain},\ }\href
  {https://doi.org/10.1103/PhysRevResearch.3.023037} {\bibfield  {journal}
  {\bibinfo  {journal} {Phys. Rev. Res.}\ }\textbf {\bibinfo {volume} {3}},\
  \bibinfo {pages} {023037} (\bibinfo {year} {2021}{\natexlab{a}})}\BibitemShut
  {NoStop}%
\bibitem [{\citenamefont {Qi}\ \emph {et~al.}(2021{\natexlab{b}})\citenamefont
  {Qi}, \citenamefont {Xing}, \citenamefont {Zhao}, \citenamefont {Liu},
  \citenamefont {Zhang}, \citenamefont {Hu},\ and\ \citenamefont
  {Wang}}]{PhysRevB.103.085129}%
  \BibitemOpen
  \bibfield  {author} {\bibinfo {author} {\bibfnamefont {L.}~\bibnamefont
  {Qi}}, \bibinfo {author} {\bibfnamefont {Y.}~\bibnamefont {Xing}}, \bibinfo
  {author} {\bibfnamefont {X.-D.}\ \bibnamefont {Zhao}}, \bibinfo {author}
  {\bibfnamefont {S.}~\bibnamefont {Liu}}, \bibinfo {author} {\bibfnamefont
  {S.}~\bibnamefont {Zhang}}, \bibinfo {author} {\bibfnamefont
  {S.}~\bibnamefont {Hu}},\ and\ \bibinfo {author} {\bibfnamefont {H.-F.}\
  \bibnamefont {Wang}},\ }\bibfield  {title} {\bibinfo {title} {Topological
  beam splitter via defect-induced edge channel in the rice-mele model},\
  }\href {https://doi.org/10.1103/PhysRevB.103.085129} {\bibfield  {journal}
  {\bibinfo  {journal} {Phys. Rev. B}\ }\textbf {\bibinfo {volume} {103}},\
  \bibinfo {pages} {085129} (\bibinfo {year} {2021}{\natexlab{b}})}\BibitemShut
  {NoStop}%
\bibitem [{\citenamefont {Brehm}\ \emph {et~al.}(2021)\citenamefont {Brehm},
  \citenamefont {Poddubny}, \citenamefont {Stehli}, \citenamefont {Wolz},
  \citenamefont {Rotzinger},\ and\ \citenamefont {Ustinov}}]{Brehm2021}%
  \BibitemOpen
  \bibfield  {author} {\bibinfo {author} {\bibfnamefont {J.~D.}\ \bibnamefont
  {Brehm}}, \bibinfo {author} {\bibfnamefont {A.~N.}\ \bibnamefont {Poddubny}},
  \bibinfo {author} {\bibfnamefont {A.}~\bibnamefont {Stehli}}, \bibinfo
  {author} {\bibfnamefont {T.}~\bibnamefont {Wolz}}, \bibinfo {author}
  {\bibfnamefont {H.}~\bibnamefont {Rotzinger}},\ and\ \bibinfo {author}
  {\bibfnamefont {A.~V.}\ \bibnamefont {Ustinov}},\ }\bibfield  {title}
  {\bibinfo {title} {{Waveguide bandgap engineering with an array of
  superconducting qubits}},\ }\href
  {https://doi.org/10.1038/s41535-021-00310-z} {\bibfield  {journal} {\bibinfo
  {journal} {Npj Quantum Mater.}\ }\textbf {\bibinfo {volume} {6}},\ \bibinfo
  {pages} {1} (\bibinfo {year} {2021})}\BibitemShut {NoStop}%
\bibitem [{\citenamefont {Zanner}\ \emph {et~al.}(2022)\citenamefont {Zanner},
  \citenamefont {Orell}, \citenamefont {Schneider}, \citenamefont {Albert},
  \citenamefont {Oleschko}, \citenamefont {Juan}, \citenamefont {Silveri},\
  and\ \citenamefont {Kirchmair}}]{Zanner2022}%
  \BibitemOpen
  \bibfield  {author} {\bibinfo {author} {\bibfnamefont {M.}~\bibnamefont
  {Zanner}}, \bibinfo {author} {\bibfnamefont {T.}~\bibnamefont {Orell}},
  \bibinfo {author} {\bibfnamefont {C.~M.}\ \bibnamefont {Schneider}}, \bibinfo
  {author} {\bibfnamefont {R.}~\bibnamefont {Albert}}, \bibinfo {author}
  {\bibfnamefont {S.}~\bibnamefont {Oleschko}}, \bibinfo {author}
  {\bibfnamefont {M.~L.}\ \bibnamefont {Juan}}, \bibinfo {author}
  {\bibfnamefont {M.}~\bibnamefont {Silveri}},\ and\ \bibinfo {author}
  {\bibfnamefont {G.}~\bibnamefont {Kirchmair}},\ }\bibfield  {title} {\bibinfo
  {title} {{Coherent control of a multi-qubit dark state in waveguide quantum
  electrodynamics}},\ }\href {https://doi.org/10.1038/s41567-022-01527-w}
  {\bibfield  {journal} {\bibinfo  {journal} {Nat. Phys}\ }\textbf {\bibinfo
  {volume} {18}},\ \bibinfo {pages} {538} (\bibinfo {year} {2022})}\BibitemShut
  {NoStop}%
\bibitem [{\citenamefont {Goban}\ \emph {et~al.}(2014)\citenamefont {Goban},
  \citenamefont {Hung}, \citenamefont {Yu}, \citenamefont {Hood}, \citenamefont
  {Muniz}, \citenamefont {Lee}, \citenamefont {Martin}, \citenamefont
  {McClung}, \citenamefont {Choi}, \citenamefont {Chang}, \citenamefont
  {Painter},\ and\ \citenamefont {Kimble}}]{goban2014atom}%
  \BibitemOpen
  \bibfield  {author} {\bibinfo {author} {\bibfnamefont {A.}~\bibnamefont
  {Goban}}, \bibinfo {author} {\bibfnamefont {C.-L.}\ \bibnamefont {Hung}},
  \bibinfo {author} {\bibfnamefont {S.-P.}\ \bibnamefont {Yu}}, \bibinfo
  {author} {\bibfnamefont {J.}~\bibnamefont {Hood}}, \bibinfo {author}
  {\bibfnamefont {J.}~\bibnamefont {Muniz}}, \bibinfo {author} {\bibfnamefont
  {J.}~\bibnamefont {Lee}}, \bibinfo {author} {\bibfnamefont {M.}~\bibnamefont
  {Martin}}, \bibinfo {author} {\bibfnamefont {A.}~\bibnamefont {McClung}},
  \bibinfo {author} {\bibfnamefont {K.}~\bibnamefont {Choi}}, \bibinfo {author}
  {\bibfnamefont {D.~E.}\ \bibnamefont {Chang}}, \bibinfo {author}
  {\bibfnamefont {O.}~\bibnamefont {Painter}},\ and\ \bibinfo {author}
  {\bibfnamefont {H.}~\bibnamefont {Kimble}},\ }\bibfield  {title} {\bibinfo
  {title} {Atom-light interactions in photonic crystals},\ }\href
  {https://doi.org/10.1038/ncomms4808} {\bibfield  {journal} {\bibinfo
  {journal} {Nat. Commun.}\ }\textbf {\bibinfo {volume} {5}},\ \bibinfo {pages}
  {3808} (\bibinfo {year} {2014})}\BibitemShut {NoStop}%
\bibitem [{\citenamefont {Goban}\ \emph {et~al.}(2015)\citenamefont {Goban},
  \citenamefont {Hung}, \citenamefont {Hood}, \citenamefont {Yu}, \citenamefont
  {Muniz}, \citenamefont {Painter},\ and\ \citenamefont
  {Kimble}}]{PhysRevLett.115.063601}%
  \BibitemOpen
  \bibfield  {author} {\bibinfo {author} {\bibfnamefont {A.}~\bibnamefont
  {Goban}}, \bibinfo {author} {\bibfnamefont {C.-L.}\ \bibnamefont {Hung}},
  \bibinfo {author} {\bibfnamefont {J.~D.}\ \bibnamefont {Hood}}, \bibinfo
  {author} {\bibfnamefont {S.-P.}\ \bibnamefont {Yu}}, \bibinfo {author}
  {\bibfnamefont {J.~A.}\ \bibnamefont {Muniz}}, \bibinfo {author}
  {\bibfnamefont {O.}~\bibnamefont {Painter}},\ and\ \bibinfo {author}
  {\bibfnamefont {H.~J.}\ \bibnamefont {Kimble}},\ }\bibfield  {title}
  {\bibinfo {title} {Superradiance for atoms trapped along a photonic crystal
  waveguide},\ }\href {https://doi.org/10.1103/PhysRevLett.115.063601}
  {\bibfield  {journal} {\bibinfo  {journal} {Phys. Rev. Lett.}\ }\textbf
  {\bibinfo {volume} {115}},\ \bibinfo {pages} {063601} (\bibinfo {year}
  {2015})}\BibitemShut {NoStop}%
\bibitem [{\citenamefont {Roy}\ \emph {et~al.}(2017)\citenamefont {Roy},
  \citenamefont {Wilson},\ and\ \citenamefont
  {Firstenberg}}]{RevModPhys.89.021001}%
  \BibitemOpen
  \bibfield  {author} {\bibinfo {author} {\bibfnamefont {D.}~\bibnamefont
  {Roy}}, \bibinfo {author} {\bibfnamefont {C.~M.}\ \bibnamefont {Wilson}},\
  and\ \bibinfo {author} {\bibfnamefont {O.}~\bibnamefont {Firstenberg}},\
  }\bibfield  {title} {\bibinfo {title} {Colloquium: Strongly interacting
  photons in one-dimensional continuum},\ }\href
  {https://doi.org/10.1103/RevModPhys.89.021001} {\bibfield  {journal}
  {\bibinfo  {journal} {Rev. Mod. Phys.}\ }\textbf {\bibinfo {volume} {89}},\
  \bibinfo {pages} {021001} (\bibinfo {year} {2017})}\BibitemShut {NoStop}%
\bibitem [{\citenamefont {Hoi}\ \emph {et~al.}(2013)\citenamefont {Hoi},
  \citenamefont {Kockum}, \citenamefont {Palomaki}, \citenamefont {Stace},
  \citenamefont {Fan}, \citenamefont {Tornberg}, \citenamefont {Sathyamoorthy},
  \citenamefont {Johansson}, \citenamefont {Delsing},\ and\ \citenamefont
  {Wilson}}]{PhysRevLett.111.053601}%
  \BibitemOpen
  \bibfield  {author} {\bibinfo {author} {\bibfnamefont {I.-C.}\ \bibnamefont
  {Hoi}}, \bibinfo {author} {\bibfnamefont {A.~F.}\ \bibnamefont {Kockum}},
  \bibinfo {author} {\bibfnamefont {T.}~\bibnamefont {Palomaki}}, \bibinfo
  {author} {\bibfnamefont {T.~M.}\ \bibnamefont {Stace}}, \bibinfo {author}
  {\bibfnamefont {B.}~\bibnamefont {Fan}}, \bibinfo {author} {\bibfnamefont
  {L.}~\bibnamefont {Tornberg}}, \bibinfo {author} {\bibfnamefont {S.~R.}\
  \bibnamefont {Sathyamoorthy}}, \bibinfo {author} {\bibfnamefont
  {G.}~\bibnamefont {Johansson}}, \bibinfo {author} {\bibfnamefont
  {P.}~\bibnamefont {Delsing}},\ and\ \bibinfo {author} {\bibfnamefont {C.~M.}\
  \bibnamefont {Wilson}},\ }\bibfield  {title} {\bibinfo {title} {Giant
  cross--kerr effect for propagating microwaves induced by an artificial
  atom},\ }\href {https://doi.org/10.1103/PhysRevLett.111.053601} {\bibfield
  {journal} {\bibinfo  {journal} {Phys. Rev. Lett.}\ }\textbf {\bibinfo
  {volume} {111}},\ \bibinfo {pages} {053601} (\bibinfo {year}
  {2013})}\BibitemShut {NoStop}%
\bibitem [{\citenamefont {Zhang}\ \emph {et~al.}(2023)\citenamefont {Zhang},
  \citenamefont {Kim}, \citenamefont {Mark}, \citenamefont {Choi},\ and\
  \citenamefont {Painter}}]{doi:10.1126/science.ade7651}%
  \BibitemOpen
  \bibfield  {author} {\bibinfo {author} {\bibfnamefont {X.}~\bibnamefont
  {Zhang}}, \bibinfo {author} {\bibfnamefont {E.}~\bibnamefont {Kim}}, \bibinfo
  {author} {\bibfnamefont {D.~K.}\ \bibnamefont {Mark}}, \bibinfo {author}
  {\bibfnamefont {S.}~\bibnamefont {Choi}},\ and\ \bibinfo {author}
  {\bibfnamefont {O.}~\bibnamefont {Painter}},\ }\bibfield  {title} {\bibinfo
  {title} {A superconducting quantum simulator based on a photonic-bandgap
  metamaterial},\ }\href {https://doi.org/10.1126/science.ade7651} {\bibfield
  {journal} {\bibinfo  {journal} {Science}\ }\textbf {\bibinfo {volume}
  {379}},\ \bibinfo {pages} {278} (\bibinfo {year} {2023})}\BibitemShut
  {NoStop}%
\bibitem [{\citenamefont {Tiranov}\ \emph {et~al.}(2023)\citenamefont
  {Tiranov}, \citenamefont {Angelopoulou}, \citenamefont {van Diepen},
  \citenamefont {Schrinski}, \citenamefont {Sandberg}, \citenamefont {Wang},
  \citenamefont {Midolo}, \citenamefont {Scholz}, \citenamefont {Wieck},
  \citenamefont {Ludwig}, \citenamefont {Sørensen},\ and\ \citenamefont
  {Lodahl}}]{doi:10.1126/science.ade9324}%
  \BibitemOpen
  \bibfield  {author} {\bibinfo {author} {\bibfnamefont {A.}~\bibnamefont
  {Tiranov}}, \bibinfo {author} {\bibfnamefont {V.}~\bibnamefont
  {Angelopoulou}}, \bibinfo {author} {\bibfnamefont {C.~J.}\ \bibnamefont {van
  Diepen}}, \bibinfo {author} {\bibfnamefont {B.}~\bibnamefont {Schrinski}},
  \bibinfo {author} {\bibfnamefont {O.~A.~D.}\ \bibnamefont {Sandberg}},
  \bibinfo {author} {\bibfnamefont {Y.}~\bibnamefont {Wang}}, \bibinfo {author}
  {\bibfnamefont {L.}~\bibnamefont {Midolo}}, \bibinfo {author} {\bibfnamefont
  {S.}~\bibnamefont {Scholz}}, \bibinfo {author} {\bibfnamefont {A.~D.}\
  \bibnamefont {Wieck}}, \bibinfo {author} {\bibfnamefont {A.}~\bibnamefont
  {Ludwig}}, \bibinfo {author} {\bibfnamefont {A.~S.}\ \bibnamefont
  {Sørensen}},\ and\ \bibinfo {author} {\bibfnamefont {P.}~\bibnamefont
  {Lodahl}},\ }\bibfield  {title} {\bibinfo {title} {Collective super- and
  subradiant dynamics between distant optical quantum emitters},\ }\href
  {https://doi.org/10.1126/science.ade9324} {\bibfield  {journal} {\bibinfo
  {journal} {Science}\ }\textbf {\bibinfo {volume} {379}},\ \bibinfo {pages}
  {389} (\bibinfo {year} {2023})}\BibitemShut {NoStop}%
\bibitem [{\citenamefont {van Loo}\ \emph {et~al.}(2013)\citenamefont {van
  Loo}, \citenamefont {Fedorov}, \citenamefont {Lalumière}, \citenamefont
  {Sanders}, \citenamefont {Blais},\ and\ \citenamefont
  {Wallraff}}]{doi:10.1126/science.1244324}%
  \BibitemOpen
  \bibfield  {author} {\bibinfo {author} {\bibfnamefont {A.~F.}\ \bibnamefont
  {van Loo}}, \bibinfo {author} {\bibfnamefont {A.}~\bibnamefont {Fedorov}},
  \bibinfo {author} {\bibfnamefont {K.}~\bibnamefont {Lalumière}}, \bibinfo
  {author} {\bibfnamefont {B.~C.}\ \bibnamefont {Sanders}}, \bibinfo {author}
  {\bibfnamefont {A.}~\bibnamefont {Blais}},\ and\ \bibinfo {author}
  {\bibfnamefont {A.}~\bibnamefont {Wallraff}},\ }\bibfield  {title} {\bibinfo
  {title} {Photon-mediated interactions between distant artificial atoms},\
  }\href {https://doi.org/10.1126/science.1244324} {\bibfield  {journal}
  {\bibinfo  {journal} {Science}\ }\textbf {\bibinfo {volume} {342}},\ \bibinfo
  {pages} {1494} (\bibinfo {year} {2013})}\BibitemShut {NoStop}%
\bibitem [{\citenamefont {Sheremet}\ \emph {et~al.}(2023)\citenamefont
  {Sheremet}, \citenamefont {Petrov}, \citenamefont {Iorsh}, \citenamefont
  {Poshakinskiy},\ and\ \citenamefont {Poddubny}}]{RevModPhys.95.015002}%
  \BibitemOpen
  \bibfield  {author} {\bibinfo {author} {\bibfnamefont {A.~S.}\ \bibnamefont
  {Sheremet}}, \bibinfo {author} {\bibfnamefont {M.~I.}\ \bibnamefont
  {Petrov}}, \bibinfo {author} {\bibfnamefont {I.~V.}\ \bibnamefont {Iorsh}},
  \bibinfo {author} {\bibfnamefont {A.~V.}\ \bibnamefont {Poshakinskiy}},\ and\
  \bibinfo {author} {\bibfnamefont {A.~N.}\ \bibnamefont {Poddubny}},\
  }\bibfield  {title} {\bibinfo {title} {Waveguide quantum electrodynamics:
  Collective radiance and photon-photon correlations},\ }\href
  {https://doi.org/10.1103/RevModPhys.95.015002} {\bibfield  {journal}
  {\bibinfo  {journal} {Rev. Mod. Phys.}\ }\textbf {\bibinfo {volume} {95}},\
  \bibinfo {pages} {015002} (\bibinfo {year} {2023})}\BibitemShut {NoStop}%
\bibitem [{\citenamefont {Douglas}\ \emph {et~al.}(2015)\citenamefont
  {Douglas}, \citenamefont {Habibian}, \citenamefont {Hung}, \citenamefont
  {Gorshkov}, \citenamefont {Kimble},\ and\ \citenamefont
  {Chang}}]{Douglas2015}%
  \BibitemOpen
  \bibfield  {author} {\bibinfo {author} {\bibfnamefont {J.~S.}\ \bibnamefont
  {Douglas}}, \bibinfo {author} {\bibfnamefont {H.}~\bibnamefont {Habibian}},
  \bibinfo {author} {\bibfnamefont {C.~L.}\ \bibnamefont {Hung}}, \bibinfo
  {author} {\bibfnamefont {A.~V.}\ \bibnamefont {Gorshkov}}, \bibinfo {author}
  {\bibfnamefont {H.~J.}\ \bibnamefont {Kimble}},\ and\ \bibinfo {author}
  {\bibfnamefont {D.~E.}\ \bibnamefont {Chang}},\ }\bibfield  {title} {\bibinfo
  {title} {{Quantum many-body models with cold atoms coupled to photonic
  crystals}},\ }\href {https://doi.org/10.1038/nphoton.2015.57} {\bibfield
  {journal} {\bibinfo  {journal} {Nat. Photonics}\ }\textbf {\bibinfo {volume}
  {9}},\ \bibinfo {pages} {326} (\bibinfo {year} {2015})}\BibitemShut {NoStop}%
\bibitem [{\citenamefont {Georgescu}\ \emph {et~al.}(2014)\citenamefont
  {Georgescu}, \citenamefont {Ashhab},\ and\ \citenamefont
  {Nori}}]{RevModPhys.86.153}%
  \BibitemOpen
  \bibfield  {author} {\bibinfo {author} {\bibfnamefont {I.~M.}\ \bibnamefont
  {Georgescu}}, \bibinfo {author} {\bibfnamefont {S.}~\bibnamefont {Ashhab}},\
  and\ \bibinfo {author} {\bibfnamefont {F.}~\bibnamefont {Nori}},\ }\bibfield
  {title} {\bibinfo {title} {Quantum simulation},\ }\href
  {https://doi.org/10.1103/RevModPhys.86.153} {\bibfield  {journal} {\bibinfo
  {journal} {Rev. Mod. Phys.}\ }\textbf {\bibinfo {volume} {86}},\ \bibinfo
  {pages} {153} (\bibinfo {year} {2014})}\BibitemShut {NoStop}%
\bibitem [{\citenamefont {Kim}\ \emph {et~al.}(2021)\citenamefont {Kim},
  \citenamefont {Zhang}, \citenamefont {Ferreira}, \citenamefont {Banker},
  \citenamefont {Iverson}, \citenamefont {Sipahigil}, \citenamefont {Bello},
  \citenamefont {Gonz\'alez-Tudela}, \citenamefont {Mirhosseini},\ and\
  \citenamefont {Painter}}]{PhysRevX.11.011015}%
  \BibitemOpen
  \bibfield  {author} {\bibinfo {author} {\bibfnamefont {E.}~\bibnamefont
  {Kim}}, \bibinfo {author} {\bibfnamefont {X.}~\bibnamefont {Zhang}}, \bibinfo
  {author} {\bibfnamefont {V.~S.}\ \bibnamefont {Ferreira}}, \bibinfo {author}
  {\bibfnamefont {J.}~\bibnamefont {Banker}}, \bibinfo {author} {\bibfnamefont
  {J.~K.}\ \bibnamefont {Iverson}}, \bibinfo {author} {\bibfnamefont
  {A.}~\bibnamefont {Sipahigil}}, \bibinfo {author} {\bibfnamefont
  {M.}~\bibnamefont {Bello}}, \bibinfo {author} {\bibfnamefont
  {A.}~\bibnamefont {Gonz\'alez-Tudela}}, \bibinfo {author} {\bibfnamefont
  {M.}~\bibnamefont {Mirhosseini}},\ and\ \bibinfo {author} {\bibfnamefont
  {O.}~\bibnamefont {Painter}},\ }\bibfield  {title} {\bibinfo {title} {Quantum
  electrodynamics in a topological waveguide},\ }\href
  {https://doi.org/10.1103/PhysRevX.11.011015} {\bibfield  {journal} {\bibinfo
  {journal} {Phys. Rev. X}\ }\textbf {\bibinfo {volume} {11}},\ \bibinfo
  {pages} {011015} (\bibinfo {year} {2021})}\BibitemShut {NoStop}%
\bibitem [{\citenamefont {Wang}\ \emph
  {et~al.}(2022{\natexlab{b}})\citenamefont {Wang}, \citenamefont {Zhao},
  \citenamefont {Chao}, \citenamefont {Peng}, \citenamefont {Yang},
  \citenamefont {Yang},\ and\ \citenamefont {Zhou}}]{Wang:22}%
  \BibitemOpen
  \bibfield  {author} {\bibinfo {author} {\bibfnamefont {D.-W.}\ \bibnamefont
  {Wang}}, \bibinfo {author} {\bibfnamefont {C.-S.}\ \bibnamefont {Zhao}},
  \bibinfo {author} {\bibfnamefont {S.-L.}\ \bibnamefont {Chao}}, \bibinfo
  {author} {\bibfnamefont {R.}~\bibnamefont {Peng}}, \bibinfo {author}
  {\bibfnamefont {J.}~\bibnamefont {Yang}}, \bibinfo {author} {\bibfnamefont
  {Z.}~\bibnamefont {Yang}},\ and\ \bibinfo {author} {\bibfnamefont
  {L.}~\bibnamefont {Zhou}},\ }\bibfield  {title} {\bibinfo {title} {Simulating
  topological phases with atom arrays in an optical waveguide},\ }\href
  {https://doi.org/10.1364/OE.472403} {\bibfield  {journal} {\bibinfo
  {journal} {Opt. Express}\ }\textbf {\bibinfo {volume} {30}},\ \bibinfo
  {pages} {42347} (\bibinfo {year} {2022}{\natexlab{b}})}\BibitemShut {NoStop}%
\bibitem [{\citenamefont {Liu}\ and\ \citenamefont {Houck}(2017)}]{Liu2017}%
  \BibitemOpen
  \bibfield  {author} {\bibinfo {author} {\bibfnamefont {Y.}~\bibnamefont
  {Liu}}\ and\ \bibinfo {author} {\bibfnamefont {A.~A.}\ \bibnamefont
  {Houck}},\ }\bibfield  {title} {\bibinfo {title} {{Quantum electrodynamics
  near a photonic bandgap}},\ }\href {https://doi.org/10.1038/nphys3834}
  {\bibfield  {journal} {\bibinfo  {journal} {Nat. Phys}\ }\textbf {\bibinfo
  {volume} {13}},\ \bibinfo {pages} {48} (\bibinfo {year} {2017})}\BibitemShut
  {NoStop}%
\bibitem [{\citenamefont {Wen}\ \emph {et~al.}(2019)\citenamefont {Wen},
  \citenamefont {Lin}, \citenamefont {Kockum}, \citenamefont {Suri},
  \citenamefont {Ian}, \citenamefont {Chen}, \citenamefont {Mao}, \citenamefont
  {Chiu}, \citenamefont {Delsing}, \citenamefont {Nori}, \citenamefont {Lin},\
  and\ \citenamefont {Hoi}}]{PhysRevLett.123.233602}%
  \BibitemOpen
  \bibfield  {author} {\bibinfo {author} {\bibfnamefont {P.~Y.}\ \bibnamefont
  {Wen}}, \bibinfo {author} {\bibfnamefont {K.-T.}\ \bibnamefont {Lin}},
  \bibinfo {author} {\bibfnamefont {A.~F.}\ \bibnamefont {Kockum}}, \bibinfo
  {author} {\bibfnamefont {B.}~\bibnamefont {Suri}}, \bibinfo {author}
  {\bibfnamefont {H.}~\bibnamefont {Ian}}, \bibinfo {author} {\bibfnamefont
  {J.~C.}\ \bibnamefont {Chen}}, \bibinfo {author} {\bibfnamefont {S.~Y.}\
  \bibnamefont {Mao}}, \bibinfo {author} {\bibfnamefont {C.~C.}\ \bibnamefont
  {Chiu}}, \bibinfo {author} {\bibfnamefont {P.}~\bibnamefont {Delsing}},
  \bibinfo {author} {\bibfnamefont {F.}~\bibnamefont {Nori}}, \bibinfo {author}
  {\bibfnamefont {G.-D.}\ \bibnamefont {Lin}},\ and\ \bibinfo {author}
  {\bibfnamefont {I.-C.}\ \bibnamefont {Hoi}},\ }\bibfield  {title} {\bibinfo
  {title} {Large collective lamb shift of two distant superconducting
  artificial atoms},\ }\href {https://doi.org/10.1103/PhysRevLett.123.233602}
  {\bibfield  {journal} {\bibinfo  {journal} {Phys. Rev. Lett.}\ }\textbf
  {\bibinfo {volume} {123}},\ \bibinfo {pages} {233602} (\bibinfo {year}
  {2019})}\BibitemShut {NoStop}%
\bibitem [{\citenamefont {John}\ and\ \citenamefont
  {Wang}(1990)}]{PhysRevLett.64.2418}%
  \BibitemOpen
  \bibfield  {author} {\bibinfo {author} {\bibfnamefont {S.}~\bibnamefont
  {John}}\ and\ \bibinfo {author} {\bibfnamefont {J.}~\bibnamefont {Wang}},\
  }\bibfield  {title} {\bibinfo {title} {Quantum electrodynamics near a
  photonic band gap: Photon bound states and dressed atoms},\ }\href
  {https://doi.org/10.1103/PhysRevLett.64.2418} {\bibfield  {journal} {\bibinfo
   {journal} {Phys. Rev. Lett.}\ }\textbf {\bibinfo {volume} {64}},\ \bibinfo
  {pages} {2418} (\bibinfo {year} {1990})}\BibitemShut {NoStop}%
\bibitem [{\citenamefont {Shi}\ \emph {et~al.}(2016)\citenamefont {Shi},
  \citenamefont {Wu}, \citenamefont {Gonz\'alez-Tudela},\ and\ \citenamefont
  {Cirac}}]{PhysRevX.6.021027}%
  \BibitemOpen
  \bibfield  {author} {\bibinfo {author} {\bibfnamefont {T.}~\bibnamefont
  {Shi}}, \bibinfo {author} {\bibfnamefont {Y.-H.}\ \bibnamefont {Wu}},
  \bibinfo {author} {\bibfnamefont {A.}~\bibnamefont {Gonz\'alez-Tudela}},\
  and\ \bibinfo {author} {\bibfnamefont {J.~I.}\ \bibnamefont {Cirac}},\
  }\bibfield  {title} {\bibinfo {title} {Bound states in boson impurity
  models},\ }\href {https://doi.org/10.1103/PhysRevX.6.021027} {\bibfield
  {journal} {\bibinfo  {journal} {Phys. Rev. X}\ }\textbf {\bibinfo {volume}
  {6}},\ \bibinfo {pages} {021027} (\bibinfo {year} {2016})}\BibitemShut
  {NoStop}%
\bibitem [{\citenamefont {Gonz\'alez-Tudela}\ and\ \citenamefont
  {Porras}(2013)}]{PhysRevLett.110.080502}%
  \BibitemOpen
  \bibfield  {author} {\bibinfo {author} {\bibfnamefont {A.}~\bibnamefont
  {Gonz\'alez-Tudela}}\ and\ \bibinfo {author} {\bibfnamefont {D.}~\bibnamefont
  {Porras}},\ }\bibfield  {title} {\bibinfo {title} {Mesoscopic entanglement
  induced by spontaneous emission in solid-state quantum optics},\ }\href
  {https://doi.org/10.1103/PhysRevLett.110.080502} {\bibfield  {journal}
  {\bibinfo  {journal} {Phys. Rev. Lett.}\ }\textbf {\bibinfo {volume} {110}},\
  \bibinfo {pages} {080502} (\bibinfo {year} {2013})}\BibitemShut {NoStop}%
\bibitem [{\citenamefont {Gonzalez-Tudela}\ \emph {et~al.}(2011)\citenamefont
  {Gonzalez-Tudela}, \citenamefont {Martin-Cano}, \citenamefont {Moreno},
  \citenamefont {Martin-Moreno}, \citenamefont {Tejedor},\ and\ \citenamefont
  {Garcia-Vidal}}]{PhysRevLett.106.020501}%
  \BibitemOpen
  \bibfield  {author} {\bibinfo {author} {\bibfnamefont {A.}~\bibnamefont
  {Gonzalez-Tudela}}, \bibinfo {author} {\bibfnamefont {D.}~\bibnamefont
  {Martin-Cano}}, \bibinfo {author} {\bibfnamefont {E.}~\bibnamefont {Moreno}},
  \bibinfo {author} {\bibfnamefont {L.}~\bibnamefont {Martin-Moreno}}, \bibinfo
  {author} {\bibfnamefont {C.}~\bibnamefont {Tejedor}},\ and\ \bibinfo {author}
  {\bibfnamefont {F.~J.}\ \bibnamefont {Garcia-Vidal}},\ }\bibfield  {title}
  {\bibinfo {title} {Entanglement of two qubits mediated by one-dimensional
  plasmonic waveguides},\ }\href
  {https://doi.org/10.1103/PhysRevLett.106.020501} {\bibfield  {journal}
  {\bibinfo  {journal} {Phys. Rev. Lett.}\ }\textbf {\bibinfo {volume} {106}},\
  \bibinfo {pages} {020501} (\bibinfo {year} {2011})}\BibitemShut {NoStop}%
\bibitem [{\citenamefont {Sayrin}\ \emph {et~al.}(2015)\citenamefont {Sayrin},
  \citenamefont {Junge}, \citenamefont {Mitsch}, \citenamefont {Albrecht},
  \citenamefont {O'Shea}, \citenamefont {Schneeweiss}, \citenamefont {Volz},\
  and\ \citenamefont {Rauschenbeutel}}]{PhysRevX.5.041036}%
  \BibitemOpen
  \bibfield  {author} {\bibinfo {author} {\bibfnamefont {C.}~\bibnamefont
  {Sayrin}}, \bibinfo {author} {\bibfnamefont {C.}~\bibnamefont {Junge}},
  \bibinfo {author} {\bibfnamefont {R.}~\bibnamefont {Mitsch}}, \bibinfo
  {author} {\bibfnamefont {B.}~\bibnamefont {Albrecht}}, \bibinfo {author}
  {\bibfnamefont {D.}~\bibnamefont {O'Shea}}, \bibinfo {author} {\bibfnamefont
  {P.}~\bibnamefont {Schneeweiss}}, \bibinfo {author} {\bibfnamefont
  {J.}~\bibnamefont {Volz}},\ and\ \bibinfo {author} {\bibfnamefont
  {A.}~\bibnamefont {Rauschenbeutel}},\ }\bibfield  {title} {\bibinfo {title}
  {Nanophotonic optical isolator controlled by the internal state of cold
  atoms},\ }\href {https://doi.org/10.1103/PhysRevX.5.041036} {\bibfield
  {journal} {\bibinfo  {journal} {Phys. Rev. X}\ }\textbf {\bibinfo {volume}
  {5}},\ \bibinfo {pages} {041036} (\bibinfo {year} {2015})}\BibitemShut
  {NoStop}%
\bibitem [{\citenamefont {Zheng}\ \emph {et~al.}(2010)\citenamefont {Zheng},
  \citenamefont {Gauthier},\ and\ \citenamefont
  {Baranger}}]{PhysRevA.82.063816}%
  \BibitemOpen
  \bibfield  {author} {\bibinfo {author} {\bibfnamefont {H.}~\bibnamefont
  {Zheng}}, \bibinfo {author} {\bibfnamefont {D.~J.}\ \bibnamefont
  {Gauthier}},\ and\ \bibinfo {author} {\bibfnamefont {H.~U.}\ \bibnamefont
  {Baranger}},\ }\bibfield  {title} {\bibinfo {title} {Waveguide qed: Many-body
  bound-state effects in coherent and fock-state scattering from a two-level
  system},\ }\href {https://doi.org/10.1103/PhysRevA.82.063816} {\bibfield
  {journal} {\bibinfo  {journal} {Phys. Rev. A}\ }\textbf {\bibinfo {volume}
  {82}},\ \bibinfo {pages} {063816} (\bibinfo {year} {2010})}\BibitemShut
  {NoStop}%
\bibitem [{\citenamefont {Bello}\ \emph {et~al.}(2019)\citenamefont {Bello},
  \citenamefont {Platero}, \citenamefont {Cirac},\ and\ \citenamefont
  {Gonz{\'{a}}lez-Tudela}}]{Bello2019}%
  \BibitemOpen
  \bibfield  {author} {\bibinfo {author} {\bibfnamefont {M.}~\bibnamefont
  {Bello}}, \bibinfo {author} {\bibfnamefont {G.}~\bibnamefont {Platero}},
  \bibinfo {author} {\bibfnamefont {J.~I.}\ \bibnamefont {Cirac}},\ and\
  \bibinfo {author} {\bibfnamefont {A.}~\bibnamefont {Gonz{\'{a}}lez-Tudela}},\
  }\bibfield  {title} {\bibinfo {title} {{Unconventional quantum optics in
  topological waveguide QED}},\ }\href {https://doi.org/10.1126/sciadv.aaw0297}
  {\bibfield  {journal} {\bibinfo  {journal} {Sci. Adv.}\ }\textbf {\bibinfo
  {volume} {5}},\ \bibinfo {pages} {1} (\bibinfo {year} {2019})},\ \Eprint
  {https://arxiv.org/abs/1811.04390} {1811.04390} \BibitemShut {NoStop}%
\bibitem [{\citenamefont {Dong}\ \emph {et~al.}(2021)\citenamefont {Dong},
  \citenamefont {Li}, \citenamefont {Liu},\ and\ \citenamefont
  {Nori}}]{PhysRevLett.126.203601}%
  \BibitemOpen
  \bibfield  {author} {\bibinfo {author} {\bibfnamefont {X.-L.}\ \bibnamefont
  {Dong}}, \bibinfo {author} {\bibfnamefont {P.-B.}\ \bibnamefont {Li}},
  \bibinfo {author} {\bibfnamefont {T.}~\bibnamefont {Liu}},\ and\ \bibinfo
  {author} {\bibfnamefont {F.}~\bibnamefont {Nori}},\ }\bibfield  {title}
  {\bibinfo {title} {Unconventional quantum sound-matter interactions in
  spin-optomechanical-crystal hybrid systems},\ }\href
  {https://doi.org/10.1103/PhysRevLett.126.203601} {\bibfield  {journal}
  {\bibinfo  {journal} {Phys. Rev. Lett.}\ }\textbf {\bibinfo {volume} {126}},\
  \bibinfo {pages} {203601} (\bibinfo {year} {2021})}\BibitemShut {NoStop}%
\bibitem [{\citenamefont {Vega}\ \emph {et~al.}(2021)\citenamefont {Vega},
  \citenamefont {Bello}, \citenamefont {Porras},\ and\ \citenamefont
  {Gonz\'alez-Tudela}}]{PhysRevA.104.053522}%
  \BibitemOpen
  \bibfield  {author} {\bibinfo {author} {\bibfnamefont {C.}~\bibnamefont
  {Vega}}, \bibinfo {author} {\bibfnamefont {M.}~\bibnamefont {Bello}},
  \bibinfo {author} {\bibfnamefont {D.}~\bibnamefont {Porras}},\ and\ \bibinfo
  {author} {\bibfnamefont {A.}~\bibnamefont {Gonz\'alez-Tudela}},\ }\bibfield
  {title} {\bibinfo {title} {Qubit-photon bound states in topological
  waveguides with long-range hoppings},\ }\href
  {https://doi.org/10.1103/PhysRevA.104.053522} {\bibfield  {journal} {\bibinfo
   {journal} {Phys. Rev. A}\ }\textbf {\bibinfo {volume} {104}},\ \bibinfo
  {pages} {053522} (\bibinfo {year} {2021})}\BibitemShut {NoStop}%
\bibitem [{\citenamefont {Dong}\ \emph {et~al.}(2022)\citenamefont {Dong},
  \citenamefont {Shen}, \citenamefont {Gao}, \citenamefont {Li}, \citenamefont
  {Gao}, \citenamefont {Li},\ and\ \citenamefont
  {Li}}]{PhysRevResearch.4.023077}%
  \BibitemOpen
  \bibfield  {author} {\bibinfo {author} {\bibfnamefont {X.-L.}\ \bibnamefont
  {Dong}}, \bibinfo {author} {\bibfnamefont {C.-P.}\ \bibnamefont {Shen}},
  \bibinfo {author} {\bibfnamefont {S.-Y.}\ \bibnamefont {Gao}}, \bibinfo
  {author} {\bibfnamefont {H.-R.}\ \bibnamefont {Li}}, \bibinfo {author}
  {\bibfnamefont {H.}~\bibnamefont {Gao}}, \bibinfo {author} {\bibfnamefont
  {F.-L.}\ \bibnamefont {Li}},\ and\ \bibinfo {author} {\bibfnamefont {P.-B.}\
  \bibnamefont {Li}},\ }\bibfield  {title} {\bibinfo {title} {Chiral
  spin-phonon bound states and spin-spin interactions with phononic lattices},\
  }\href {https://doi.org/10.1103/PhysRevResearch.4.023077} {\bibfield
  {journal} {\bibinfo  {journal} {Phys. Rev. Res.}\ }\textbf {\bibinfo {volume}
  {4}},\ \bibinfo {pages} {023077} (\bibinfo {year} {2022})}\BibitemShut
  {NoStop}%
\bibitem [{\citenamefont {Ringel}\ \emph {et~al.}(2014)\citenamefont {Ringel},
  \citenamefont {Pletyukhov},\ and\ \citenamefont {Gritsev}}]{Ringel_2014}%
  \BibitemOpen
  \bibfield  {author} {\bibinfo {author} {\bibfnamefont {M.}~\bibnamefont
  {Ringel}}, \bibinfo {author} {\bibfnamefont {M.}~\bibnamefont {Pletyukhov}},\
  and\ \bibinfo {author} {\bibfnamefont {V.}~\bibnamefont {Gritsev}},\
  }\bibfield  {title} {\bibinfo {title} {Topologically protected strongly
  correlated states of photons},\ }\href
  {https://doi.org/10.1088/1367-2630/16/11/113030} {\bibfield  {journal}
  {\bibinfo  {journal} {New J. Phys.}\ }\textbf {\bibinfo {volume} {16}},\
  \bibinfo {pages} {113030} (\bibinfo {year} {2014})}\BibitemShut {NoStop}%
\bibitem [{\citenamefont {Ren}\ \emph {et~al.}(2022)\citenamefont {Ren},
  \citenamefont {Xie}, \citenamefont {Li}, \citenamefont {Ma},\ and\
  \citenamefont {Li}}]{PhysRevB.105.094422}%
  \BibitemOpen
  \bibfield  {author} {\bibinfo {author} {\bibfnamefont {Y.-l.}\ \bibnamefont
  {Ren}}, \bibinfo {author} {\bibfnamefont {J.-k.}\ \bibnamefont {Xie}},
  \bibinfo {author} {\bibfnamefont {X.-k.}\ \bibnamefont {Li}}, \bibinfo
  {author} {\bibfnamefont {S.-l.}\ \bibnamefont {Ma}},\ and\ \bibinfo {author}
  {\bibfnamefont {F.-l.}\ \bibnamefont {Li}},\ }\bibfield  {title} {\bibinfo
  {title} {Long-range generation of a magnon-magnon entangled state},\ }\href
  {https://doi.org/10.1103/PhysRevB.105.094422} {\bibfield  {journal} {\bibinfo
   {journal} {Phys. Rev. B}\ }\textbf {\bibinfo {volume} {105}},\ \bibinfo
  {pages} {094422} (\bibinfo {year} {2022})}\BibitemShut {NoStop}%
\bibitem [{\citenamefont {Cheng}\ \emph {et~al.}(2022)\citenamefont {Cheng},
  \citenamefont {Wang},\ and\ \citenamefont {Liu}}]{PhysRevA.106.033522}%
  \BibitemOpen
  \bibfield  {author} {\bibinfo {author} {\bibfnamefont {W.}~\bibnamefont
  {Cheng}}, \bibinfo {author} {\bibfnamefont {Z.}~\bibnamefont {Wang}},\ and\
  \bibinfo {author} {\bibfnamefont {Y.-x.}\ \bibnamefont {Liu}},\ }\bibfield
  {title} {\bibinfo {title} {Topology and retardation effect of a giant atom in
  a topological waveguide},\ }\href
  {https://doi.org/10.1103/PhysRevA.106.033522} {\bibfield  {journal} {\bibinfo
   {journal} {Phys. Rev. A}\ }\textbf {\bibinfo {volume} {106}},\ \bibinfo
  {pages} {033522} (\bibinfo {year} {2022})}\BibitemShut {NoStop}%
\bibitem [{\citenamefont {Delplace}\ \emph {et~al.}(2011)\citenamefont
  {Delplace}, \citenamefont {Ullmo},\ and\ \citenamefont
  {Montambaux}}]{PhysRevB.84.195452}%
  \BibitemOpen
  \bibfield  {author} {\bibinfo {author} {\bibfnamefont {P.}~\bibnamefont
  {Delplace}}, \bibinfo {author} {\bibfnamefont {D.}~\bibnamefont {Ullmo}},\
  and\ \bibinfo {author} {\bibfnamefont {G.}~\bibnamefont {Montambaux}},\
  }\bibfield  {title} {\bibinfo {title} {Zak phase and the existence of edge
  states in graphene},\ }\href {https://doi.org/10.1103/PhysRevB.84.195452}
  {\bibfield  {journal} {\bibinfo  {journal} {Phys. Rev. B}\ }\textbf {\bibinfo
  {volume} {84}},\ \bibinfo {pages} {195452} (\bibinfo {year}
  {2011})}\BibitemShut {NoStop}%
\bibitem [{\citenamefont {Zak}(1989)}]{PhysRevLett.62.2747}%
  \BibitemOpen
  \bibfield  {author} {\bibinfo {author} {\bibfnamefont {J.}~\bibnamefont
  {Zak}},\ }\bibfield  {title} {\bibinfo {title} {Berry's phase for energy
  bands in solids},\ }\href {https://doi.org/10.1103/PhysRevLett.62.2747}
  {\bibfield  {journal} {\bibinfo  {journal} {Phys. Rev. Lett.}\ }\textbf
  {\bibinfo {volume} {62}},\ \bibinfo {pages} {2747} (\bibinfo {year}
  {1989})}\BibitemShut {NoStop}%
\bibitem [{\citenamefont {Longhi}(2019)}]{PhysRevB.99.155150}%
  \BibitemOpen
  \bibfield  {author} {\bibinfo {author} {\bibfnamefont {S.}~\bibnamefont
  {Longhi}},\ }\bibfield  {title} {\bibinfo {title} {Topological pumping of
  edge states via adiabatic passage},\ }\href
  {https://doi.org/10.1103/PhysRevB.99.155150} {\bibfield  {journal} {\bibinfo
  {journal} {Phys. Rev. B}\ }\textbf {\bibinfo {volume} {99}},\ \bibinfo
  {pages} {155150} (\bibinfo {year} {2019})}\BibitemShut {NoStop}%
\bibitem [{\citenamefont {Blais}\ \emph {et~al.}(2021)\citenamefont {Blais},
  \citenamefont {Grimsmo}, \citenamefont {Girvin},\ and\ \citenamefont
  {Wallraff}}]{RevModPhys.93.025005}%
  \BibitemOpen
  \bibfield  {author} {\bibinfo {author} {\bibfnamefont {A.}~\bibnamefont
  {Blais}}, \bibinfo {author} {\bibfnamefont {A.~L.}\ \bibnamefont {Grimsmo}},
  \bibinfo {author} {\bibfnamefont {S.~M.}\ \bibnamefont {Girvin}},\ and\
  \bibinfo {author} {\bibfnamefont {A.}~\bibnamefont {Wallraff}},\ }\bibfield
  {title} {\bibinfo {title} {Circuit quantum electrodynamics},\ }\href
  {https://doi.org/10.1103/RevModPhys.93.025005} {\bibfield  {journal}
  {\bibinfo  {journal} {Rev. Mod. Phys.}\ }\textbf {\bibinfo {volume} {93}},\
  \bibinfo {pages} {025005} (\bibinfo {year} {2021})}\BibitemShut {NoStop}%
\bibitem [{\citenamefont {Majer}\ \emph {et~al.}(2005)\citenamefont {Majer},
  \citenamefont {Paauw}, \citenamefont {ter Haar}, \citenamefont {Harmans},\
  and\ \citenamefont {Mooij}}]{PhysRevLett.94.090501}%
  \BibitemOpen
  \bibfield  {author} {\bibinfo {author} {\bibfnamefont {J.~B.}\ \bibnamefont
  {Majer}}, \bibinfo {author} {\bibfnamefont {F.~G.}\ \bibnamefont {Paauw}},
  \bibinfo {author} {\bibfnamefont {A.~C.~J.}\ \bibnamefont {ter Haar}},
  \bibinfo {author} {\bibfnamefont {C.~J. P.~M.}\ \bibnamefont {Harmans}},\
  and\ \bibinfo {author} {\bibfnamefont {J.~E.}\ \bibnamefont {Mooij}},\
  }\bibfield  {title} {\bibinfo {title} {Spectroscopy on two coupled
  superconducting flux qubits},\ }\href
  {https://doi.org/10.1103/PhysRevLett.94.090501} {\bibfield  {journal}
  {\bibinfo  {journal} {Phys. Rev. Lett.}\ }\textbf {\bibinfo {volume} {94}},\
  \bibinfo {pages} {090501} (\bibinfo {year} {2005})}\BibitemShut {NoStop}%
\bibitem [{\citenamefont {Vion}\ \emph {et~al.}(2002)\citenamefont {Vion},
  \citenamefont {Aassime}, \citenamefont {Cottet}, \citenamefont {Joyez},
  \citenamefont {Pothier}, \citenamefont {Urbina}, \citenamefont {Esteve},\
  and\ \citenamefont {Devoret}}]{doi:10.1126/science.1069372}%
  \BibitemOpen
  \bibfield  {author} {\bibinfo {author} {\bibfnamefont {D.}~\bibnamefont
  {Vion}}, \bibinfo {author} {\bibfnamefont {A.}~\bibnamefont {Aassime}},
  \bibinfo {author} {\bibfnamefont {A.}~\bibnamefont {Cottet}}, \bibinfo
  {author} {\bibfnamefont {P.}~\bibnamefont {Joyez}}, \bibinfo {author}
  {\bibfnamefont {H.}~\bibnamefont {Pothier}}, \bibinfo {author} {\bibfnamefont
  {C.}~\bibnamefont {Urbina}}, \bibinfo {author} {\bibfnamefont
  {D.}~\bibnamefont {Esteve}},\ and\ \bibinfo {author} {\bibfnamefont {M.~H.}\
  \bibnamefont {Devoret}},\ }\bibfield  {title} {\bibinfo {title} {Manipulating
  the quantum state of an electrical circuit},\ }\href
  {https://www.science.org/doi/abs/10.1126/science.1069372} {\bibfield
  {journal} {\bibinfo  {journal} {Science}\ }\textbf {\bibinfo {volume}
  {296}},\ \bibinfo {pages} {886} (\bibinfo {year} {2002})}\BibitemShut
  {NoStop}%
\bibitem [{\citenamefont {Buluta}\ and\ \citenamefont
  {Nori}(2009)}]{doi:10.1126/science.1177838}%
  \BibitemOpen
  \bibfield  {author} {\bibinfo {author} {\bibfnamefont {I.}~\bibnamefont
  {Buluta}}\ and\ \bibinfo {author} {\bibfnamefont {F.}~\bibnamefont {Nori}},\
  }\bibfield  {title} {\bibinfo {title} {Quantum simulators},\ }\href
  {https://www.science.org/doi/abs/10.1126/science.1177838} {\bibfield
  {journal} {\bibinfo  {journal} {Science}\ }\textbf {\bibinfo {volume}
  {326}},\ \bibinfo {pages} {108} (\bibinfo {year} {2009})}\BibitemShut
  {NoStop}%
\bibitem [{\citenamefont {Fitzpatrick}\ \emph {et~al.}(2017)\citenamefont
  {Fitzpatrick}, \citenamefont {Sundaresan}, \citenamefont {Li}, \citenamefont
  {Koch},\ and\ \citenamefont {Houck}}]{PhysRevX.7.011016}%
  \BibitemOpen
  \bibfield  {author} {\bibinfo {author} {\bibfnamefont {M.}~\bibnamefont
  {Fitzpatrick}}, \bibinfo {author} {\bibfnamefont {N.~M.}\ \bibnamefont
  {Sundaresan}}, \bibinfo {author} {\bibfnamefont {A.~C.~Y.}\ \bibnamefont
  {Li}}, \bibinfo {author} {\bibfnamefont {J.}~\bibnamefont {Koch}},\ and\
  \bibinfo {author} {\bibfnamefont {A.~A.}\ \bibnamefont {Houck}},\ }\bibfield
  {title} {\bibinfo {title} {Observation of a dissipative phase transition in a
  one-dimensional circuit qed lattice},\ }\href
  {https://doi.org/10.1103/PhysRevX.7.011016} {\bibfield  {journal} {\bibinfo
  {journal} {Phys. Rev. X}\ }\textbf {\bibinfo {volume} {7}},\ \bibinfo {pages}
  {011016} (\bibinfo {year} {2017})}\BibitemShut {NoStop}%
\bibitem [{\citenamefont {Noh}\ and\ \citenamefont
  {Angelakis}(2016)}]{Noh2017}%
  \BibitemOpen
  \bibfield  {author} {\bibinfo {author} {\bibfnamefont {C.}~\bibnamefont
  {Noh}}\ and\ \bibinfo {author} {\bibfnamefont {D.~G.}\ \bibnamefont
  {Angelakis}},\ }\bibfield  {title} {\bibinfo {title} {Quantum simulations and
  many-body physics with light},\ }\href
  {https://doi.org/10.1088/0034-4885/80/1/016401} {\bibfield  {journal}
  {\bibinfo  {journal} {Rep. Prog. Phys.}\ }\textbf {\bibinfo {volume} {80}},\
  \bibinfo {pages} {016401} (\bibinfo {year} {2016})}\BibitemShut {NoStop}%
\bibitem [{\citenamefont {Daley}\ \emph {et~al.}(2022)\citenamefont {Daley},
  \citenamefont {Bloch}, \citenamefont {Kokail}, \citenamefont {Flannigan},
  \citenamefont {Pearson}, \citenamefont {Troyer},\ and\ \citenamefont
  {Zoller}}]{Daley2022}%
  \BibitemOpen
  \bibfield  {author} {\bibinfo {author} {\bibfnamefont {A.~J.}\ \bibnamefont
  {Daley}}, \bibinfo {author} {\bibfnamefont {I.}~\bibnamefont {Bloch}},
  \bibinfo {author} {\bibfnamefont {C.}~\bibnamefont {Kokail}}, \bibinfo
  {author} {\bibfnamefont {S.}~\bibnamefont {Flannigan}}, \bibinfo {author}
  {\bibfnamefont {N.}~\bibnamefont {Pearson}}, \bibinfo {author} {\bibfnamefont
  {M.}~\bibnamefont {Troyer}},\ and\ \bibinfo {author} {\bibfnamefont
  {P.}~\bibnamefont {Zoller}},\ }\bibfield  {title} {\bibinfo {title}
  {{Practical quantum advantage in quantum simulation}},\ }\href
  {https://doi.org/10.1038/s41586-022-04940-6} {\bibfield  {journal} {\bibinfo
  {journal} {Nature}\ }\textbf {\bibinfo {volume} {607}},\ \bibinfo {pages}
  {667} (\bibinfo {year} {2022})}\BibitemShut {NoStop}%
\bibitem [{\citenamefont {Blais}\ \emph {et~al.}(2007)\citenamefont {Blais},
  \citenamefont {Gambetta}, \citenamefont {Wallraff}, \citenamefont {Schuster},
  \citenamefont {Girvin}, \citenamefont {Devoret},\ and\ \citenamefont
  {Schoelkopf}}]{PhysRevA.75.032329}%
  \BibitemOpen
  \bibfield  {author} {\bibinfo {author} {\bibfnamefont {A.}~\bibnamefont
  {Blais}}, \bibinfo {author} {\bibfnamefont {J.}~\bibnamefont {Gambetta}},
  \bibinfo {author} {\bibfnamefont {A.}~\bibnamefont {Wallraff}}, \bibinfo
  {author} {\bibfnamefont {D.~I.}\ \bibnamefont {Schuster}}, \bibinfo {author}
  {\bibfnamefont {S.~M.}\ \bibnamefont {Girvin}}, \bibinfo {author}
  {\bibfnamefont {M.~H.}\ \bibnamefont {Devoret}},\ and\ \bibinfo {author}
  {\bibfnamefont {R.~J.}\ \bibnamefont {Schoelkopf}},\ }\bibfield  {title}
  {\bibinfo {title} {Quantum-information processing with circuit quantum
  electrodynamics},\ }\href {https://doi.org/10.1103/PhysRevA.75.032329}
  {\bibfield  {journal} {\bibinfo  {journal} {Phys. Rev. A}\ }\textbf {\bibinfo
  {volume} {75}},\ \bibinfo {pages} {032329} (\bibinfo {year}
  {2007})}\BibitemShut {NoStop}%
\bibitem [{\citenamefont {Mirhosseini}\ \emph {et~al.}(2019)\citenamefont
  {Mirhosseini}, \citenamefont {Kim}, \citenamefont {Zhang}, \citenamefont
  {Sipahigil}, \citenamefont {Dieterle}, \citenamefont {Keller}, \citenamefont
  {Asenjo-Garcia}, \citenamefont {Chang},\ and\ \citenamefont
  {Painter}}]{Mirhosseini2019}%
  \BibitemOpen
  \bibfield  {author} {\bibinfo {author} {\bibfnamefont {M.}~\bibnamefont
  {Mirhosseini}}, \bibinfo {author} {\bibfnamefont {E.}~\bibnamefont {Kim}},
  \bibinfo {author} {\bibfnamefont {X.}~\bibnamefont {Zhang}}, \bibinfo
  {author} {\bibfnamefont {A.}~\bibnamefont {Sipahigil}}, \bibinfo {author}
  {\bibfnamefont {P.~B.}\ \bibnamefont {Dieterle}}, \bibinfo {author}
  {\bibfnamefont {A.~J.}\ \bibnamefont {Keller}}, \bibinfo {author}
  {\bibfnamefont {A.}~\bibnamefont {Asenjo-Garcia}}, \bibinfo {author}
  {\bibfnamefont {D.~E.}\ \bibnamefont {Chang}},\ and\ \bibinfo {author}
  {\bibfnamefont {O.}~\bibnamefont {Painter}},\ }\bibfield  {title} {\bibinfo
  {title} {{Cavity quantum electrodynamics with atom-like mirrors}},\ }\href
  {https://doi.org/10.1038/s41586-019-1196-1} {\bibfield  {journal} {\bibinfo
  {journal} {Nature}\ }\textbf {\bibinfo {volume} {569}},\ \bibinfo {pages}
  {692} (\bibinfo {year} {2019})}\BibitemShut {NoStop}%
\bibitem [{\citenamefont {Nie}\ and\ \citenamefont
  {Liu}(2020)}]{PhysRevResearch.2.012076}%
  \BibitemOpen
  \bibfield  {author} {\bibinfo {author} {\bibfnamefont {W.}~\bibnamefont
  {Nie}}\ and\ \bibinfo {author} {\bibfnamefont {Y.-x.}\ \bibnamefont {Liu}},\
  }\bibfield  {title} {\bibinfo {title} {Bandgap-assisted quantum control of
  topological edge states in a cavity},\ }\href
  {https://doi.org/10.1103/PhysRevResearch.2.012076} {\bibfield  {journal}
  {\bibinfo  {journal} {Phys. Rev. Res.}\ }\textbf {\bibinfo {volume} {2}},\
  \bibinfo {pages} {012076} (\bibinfo {year} {2020})}\BibitemShut {NoStop}%
\bibitem [{\citenamefont {Sundaresan}\ \emph {et~al.}(2019)\citenamefont
  {Sundaresan}, \citenamefont {Lundgren}, \citenamefont {Zhu}, \citenamefont
  {Gorshkov},\ and\ \citenamefont {Houck}}]{PhysRevX.9.011021}%
  \BibitemOpen
  \bibfield  {author} {\bibinfo {author} {\bibfnamefont {N.~M.}\ \bibnamefont
  {Sundaresan}}, \bibinfo {author} {\bibfnamefont {R.}~\bibnamefont
  {Lundgren}}, \bibinfo {author} {\bibfnamefont {G.}~\bibnamefont {Zhu}},
  \bibinfo {author} {\bibfnamefont {A.~V.}\ \bibnamefont {Gorshkov}},\ and\
  \bibinfo {author} {\bibfnamefont {A.~A.}\ \bibnamefont {Houck}},\ }\bibfield
  {title} {\bibinfo {title} {Interacting qubit-photon bound states with
  superconducting circuits},\ }\href
  {https://doi.org/10.1103/PhysRevX.9.011021} {\bibfield  {journal} {\bibinfo
  {journal} {Phys. Rev. X}\ }\textbf {\bibinfo {volume} {9}},\ \bibinfo {pages}
  {011021} (\bibinfo {year} {2019})}\BibitemShut {NoStop}%
\bibitem [{\citenamefont {Leonforte}\ \emph {et~al.}(2021)\citenamefont
  {Leonforte}, \citenamefont {Carollo},\ and\ \citenamefont
  {Ciccarello}}]{PhysRevLett.126.063601}%
  \BibitemOpen
  \bibfield  {author} {\bibinfo {author} {\bibfnamefont {L.}~\bibnamefont
  {Leonforte}}, \bibinfo {author} {\bibfnamefont {A.}~\bibnamefont {Carollo}},\
  and\ \bibinfo {author} {\bibfnamefont {F.}~\bibnamefont {Ciccarello}},\
  }\bibfield  {title} {\bibinfo {title} {Vacancy-like dressed states in
  topological waveguide qed},\ }\href
  {https://doi.org/10.1103/PhysRevLett.126.063601} {\bibfield  {journal}
  {\bibinfo  {journal} {Phys. Rev. Lett.}\ }\textbf {\bibinfo {volume} {126}},\
  \bibinfo {pages} {063601} (\bibinfo {year} {2021})}\BibitemShut {NoStop}%
\bibitem [{\citenamefont {Scigliuzzo}\ \emph {et~al.}(2022)\citenamefont
  {Scigliuzzo}, \citenamefont {Calaj{\`{o}}}, \citenamefont {Ciccarello},
  \citenamefont {{Perez Lozano}}, \citenamefont {Bengtsson}, \citenamefont
  {Scarlino}, \citenamefont {Wallraff}, \citenamefont {Chang}, \citenamefont
  {Delsing},\ and\ \citenamefont {Gasparinetti}}]{Scigliuzzo2022}%
  \BibitemOpen
  \bibfield  {author} {\bibinfo {author} {\bibfnamefont {M.}~\bibnamefont
  {Scigliuzzo}}, \bibinfo {author} {\bibfnamefont {G.}~\bibnamefont
  {Calaj{\`{o}}}}, \bibinfo {author} {\bibfnamefont {F.}~\bibnamefont
  {Ciccarello}}, \bibinfo {author} {\bibfnamefont {D.}~\bibnamefont {{Perez
  Lozano}}}, \bibinfo {author} {\bibfnamefont {A.}~\bibnamefont {Bengtsson}},
  \bibinfo {author} {\bibfnamefont {P.}~\bibnamefont {Scarlino}}, \bibinfo
  {author} {\bibfnamefont {A.}~\bibnamefont {Wallraff}}, \bibinfo {author}
  {\bibfnamefont {D.}~\bibnamefont {Chang}}, \bibinfo {author} {\bibfnamefont
  {P.}~\bibnamefont {Delsing}},\ and\ \bibinfo {author} {\bibfnamefont
  {S.}~\bibnamefont {Gasparinetti}},\ }\bibfield  {title} {\bibinfo {title}
  {{Controlling Atom-Photon Bound States in an Array of Josephson-Junction
  Resonators}},\ }\href {https://doi.org/10.1103/PhysRevX.12.031036} {\bibfield
   {journal} {\bibinfo  {journal} {Phys. Rev. X}\ }\textbf {\bibinfo {volume}
  {12}},\ \bibinfo {pages} {31036} (\bibinfo {year} {2022})}\BibitemShut
  {NoStop}%
\bibitem [{\citenamefont {Houck}\ \emph {et~al.}(2012)\citenamefont {Houck},
  \citenamefont {T{\"{u}}reci},\ and\ \citenamefont {Koch}}]{Houck2012}%
  \BibitemOpen
  \bibfield  {author} {\bibinfo {author} {\bibfnamefont {A.~A.}\ \bibnamefont
  {Houck}}, \bibinfo {author} {\bibfnamefont {H.~E.}\ \bibnamefont
  {T{\"{u}}reci}},\ and\ \bibinfo {author} {\bibfnamefont {J.}~\bibnamefont
  {Koch}},\ }\bibfield  {title} {\bibinfo {title} {{On-chip quantum simulation
  with superconducting circuits}},\ }\href {https://doi.org/10.1038/nphys2251}
  {\bibfield  {journal} {\bibinfo  {journal} {Nat. Phys.}\ }\textbf {\bibinfo
  {volume} {8}},\ \bibinfo {pages} {292} (\bibinfo {year} {2012})}\BibitemShut
  {NoStop}%
\bibitem [{\citenamefont {Schmidt}\ and\ \citenamefont
  {Koch}(2013)}]{Schmidt2013}%
  \BibitemOpen
  \bibfield  {author} {\bibinfo {author} {\bibfnamefont {S.}~\bibnamefont
  {Schmidt}}\ and\ \bibinfo {author} {\bibfnamefont {J.}~\bibnamefont {Koch}},\
  }\bibfield  {title} {\bibinfo {title} {{Circuit QED lattices: Towards quantum
  simulation with superconducting circuits}},\ }\href
  {https://doi.org/10.1002/andp.201200261} {\bibfield  {journal} {\bibinfo
  {journal} {Ann. Phys.}\ }\textbf {\bibinfo {volume} {525}},\ \bibinfo {pages}
  {395} (\bibinfo {year} {2013})}\BibitemShut {NoStop}%
\bibitem [{\citenamefont {Guo}\ \emph {et~al.}(2017)\citenamefont {Guo},
  \citenamefont {Grimsmo}, \citenamefont {Kockum}, \citenamefont {Pletyukhov},\
  and\ \citenamefont {Johansson}}]{PhysRevA.95.053821}%
  \BibitemOpen
  \bibfield  {author} {\bibinfo {author} {\bibfnamefont {L.}~\bibnamefont
  {Guo}}, \bibinfo {author} {\bibfnamefont {A.}~\bibnamefont {Grimsmo}},
  \bibinfo {author} {\bibfnamefont {A.~F.}\ \bibnamefont {Kockum}}, \bibinfo
  {author} {\bibfnamefont {M.}~\bibnamefont {Pletyukhov}},\ and\ \bibinfo
  {author} {\bibfnamefont {G.}~\bibnamefont {Johansson}},\ }\bibfield  {title}
  {\bibinfo {title} {Giant acoustic atom: A single quantum system with a
  deterministic time delay},\ }\href
  {https://doi.org/10.1103/PhysRevA.95.053821} {\bibfield  {journal} {\bibinfo
  {journal} {Phys. Rev. A}\ }\textbf {\bibinfo {volume} {95}},\ \bibinfo
  {pages} {053821} (\bibinfo {year} {2017})}\BibitemShut {NoStop}%
\bibitem [{\citenamefont {Kannan}\ \emph {et~al.}(2020)\citenamefont {Kannan},
  \citenamefont {Ruckriegel}, \citenamefont {Campbell}, \citenamefont {{Frisk
  Kockum}}, \citenamefont {Braum{\"{u}}ller}, \citenamefont {Kim},
  \citenamefont {Kjaergaard}, \citenamefont {Krantz}, \citenamefont {Melville},
  \citenamefont {Niedzielski}, \citenamefont {Veps{\"{a}}l{\"{a}}inen},
  \citenamefont {Winik}, \citenamefont {Yoder}, \citenamefont {Nori},
  \citenamefont {Orlando}, \citenamefont {Gustavsson},\ and\ \citenamefont
  {Oliver}}]{Kannan2020}%
  \BibitemOpen
  \bibfield  {author} {\bibinfo {author} {\bibfnamefont {B.}~\bibnamefont
  {Kannan}}, \bibinfo {author} {\bibfnamefont {M.~J.}\ \bibnamefont
  {Ruckriegel}}, \bibinfo {author} {\bibfnamefont {D.~L.}\ \bibnamefont
  {Campbell}}, \bibinfo {author} {\bibfnamefont {A.}~\bibnamefont {{Frisk
  Kockum}}}, \bibinfo {author} {\bibfnamefont {J.}~\bibnamefont
  {Braum{\"{u}}ller}}, \bibinfo {author} {\bibfnamefont {D.~K.}\ \bibnamefont
  {Kim}}, \bibinfo {author} {\bibfnamefont {M.}~\bibnamefont {Kjaergaard}},
  \bibinfo {author} {\bibfnamefont {P.}~\bibnamefont {Krantz}}, \bibinfo
  {author} {\bibfnamefont {A.}~\bibnamefont {Melville}}, \bibinfo {author}
  {\bibfnamefont {B.~M.}\ \bibnamefont {Niedzielski}}, \bibinfo {author}
  {\bibfnamefont {A.}~\bibnamefont {Veps{\"{a}}l{\"{a}}inen}}, \bibinfo
  {author} {\bibfnamefont {R.}~\bibnamefont {Winik}}, \bibinfo {author}
  {\bibfnamefont {J.~L.}\ \bibnamefont {Yoder}}, \bibinfo {author}
  {\bibfnamefont {F.}~\bibnamefont {Nori}}, \bibinfo {author} {\bibfnamefont
  {T.~P.}\ \bibnamefont {Orlando}}, \bibinfo {author} {\bibfnamefont
  {S.}~\bibnamefont {Gustavsson}},\ and\ \bibinfo {author} {\bibfnamefont
  {W.~D.}\ \bibnamefont {Oliver}},\ }\bibfield  {title} {\bibinfo {title}
  {{Waveguide quantum electrodynamics with superconducting artificial giant
  atoms}},\ }\href {https://doi.org/10.1038/s41586-020-2529-9} {\bibfield
  {journal} {\bibinfo  {journal} {Nature}\ }\textbf {\bibinfo {volume} {583}},\
  \bibinfo {pages} {775} (\bibinfo {year} {2020})}\BibitemShut {NoStop}%
\end{thebibliography}%
\end{document}